
\input amstex
\input amsppt.sty

\TagsOnRight \NoBlackBoxes

\define\Y{\Bbb Y}
\define\Z{\Bbb Z}
\define\C{\Bbb C}
\define\R{\Bbb R}
\define\T{\Bbb T}

\define\al{\alpha}

\define\Ga{\Gamma}
\define\de{\delta}
\define\La{\Lambda}
\define\la{{\lambda}}

\define\th{{\theta}}
\define\epsi{\varepsilon}

\define\wt{\widetilde}

\define\tp{\widetilde p}
\define\tq{\widetilde q}

\def\const{\operatorname{const}}
\define\sgn{\operatorname{sgn}}
\define\Mat{\operatorname{Mat}}
\define\tr{\operatorname{tr}}
\define\Prob{\operatorname{Prob}}
\define\tail{{\operatorname{tail}}}
\define\gam{{\operatorname{gamma}}}
\define\ps{{\operatorname{psi}}}
\define\SGN{\operatorname{SGN}}
\define\Dim{\operatorname{Dim}}
\define\ESF{\operatorname{ESF}}

\define\unX{\underline X}
\define\unK{\underline K}
\define\unP{\underline P}

\define\tht{\thetag}

\define\s{\Sigma}

\define\wh{\widehat}

\topmatter

\title Random partitions and the Gamma kernel
\endtitle
\author
Alexei Borodin and Grigori Olshanski
\endauthor

\abstract We study the asymptotics of certain measures on
partitions (the so-called z-measures and their relatives) in two
different regimes: near the diagonal of the corresponding Young
diagram and in the intermediate zone between the diagonal and the
edge of the Young diagram. We prove that in both cases the limit
correlation functions have determinantal form with a correlation
kernel which depends on two real parameters. In the first case the
correlation kernel is discrete, and it has a simple expression in
terms of the gamma functions. In the second case the correlation
kernel is continuous and translationally invariant, and it can be
a written as a ratio of two suitably scaled hyperbolic sines.

\endabstract

\thanks This research was partially conducted during the period
one of the authors (A.B.) served as a Clay Mathematics Institute
Long--Term Prize Fellow.
\endthanks
\toc \widestnumber\head{???}

\head 0. Introduction \endhead \head 1. Definition of the
z-measures \endhead \head 2. The hypergeometric and gamma kernels
(first form)
\endhead
\head 3. The hypergeometric and gamma kernels (second form)
\endhead
\head 4. The relation between two forms of kernels
\endhead
\head 5. The projection property \endhead \head 6. The tail kernel
\endhead \head 7. ZW--measures on signatures \endhead \head 8.
Z--measures on nonnegative signatures
\endhead
 \head {} References \endhead
\endtoc

\endtopmatter

\document

\head 0. Introduction \endhead

In recent years there has been a lot of interest in understanding
the ``random matrix type'' limit behavior of different measures on
partitions as the size of partitions goes to infinity. The most
known result is the Baik-Deift-Johansson theorem \cite{BDJ} that
claims that the limit distribution of the (centered and scaled)
first part of the random partitions distributed according to the
so-called Plancherel measure is just the same as that of the
largest eigenvalue of random Hermitian matrices from the Gaussian
Unitary Ensemble.

The goal of this paper is to study the asymptotic behavior of the
so-called z-measures and their relatives. The asymptotics of the
largest parts of partitions distributed according to such measures
has a representation theoretic meaning: it encodes the spectral
decomposition of generalized regular representations of certain
groups into irreducibles. We have computed this asymptotics in the
cases of the infinite symmetric group and the
infinite--dimensional unitary group in our previous work, see
\cite{BO2}, \cite{BO4}. The main result of this paper is a
complete description of the limit behavior of these measures near
the diagonal (smallest Frobenius coordinates) and in the
intermediate zone between the diagonal and the edge of the
partition (Frobenius coordinates of intermediate growth).

A more detailed description of the content of the paper follows.

\subhead The z--measures \endsubhead Let $\Y_n$ denote the set of
partitions of a natural number $n$ and
$\Y=\Y_0\sqcup\Y_1\sqcup\Y_2\sqcup\dots$ be the set of all
partitions. We identify partitions and Young diagrams. We consider
a Hilbert space $H$ together with a distinguished orthonormal
basis $\{\chi_\la\}$ parameterized by $\la\in\Y$. The basis
elements may be identified with irreducible characters of
symmetric groups of arbitrary degree. Next, we construct a family
of vectors $f_{z,\xi}$ in $H$, indexed by couples
$(z,\xi)\in\C\times(0,1)$. Set
$$
M_{z,z',\xi}(\la) =\frac{(f_{z,\xi},\chi_\la)(\chi_\la,f_{\overline{z'},\xi})}
{(f_{z,\xi},f_{\overline{z'},\xi})}
$$
where $z\in\C$, $z'\in\C$, $0<\xi<1$, and $(\,\cdot\,,\,\cdot\,)$ denotes the
inner product in $H$. It turns out that
$(f_{z,\xi},f_{\overline{z'},\xi})\ne0$, so that the above expression makes
sense. Clearly,
$$
\sum_{\la\in\Y}M_{z,z',\xi}(\la)=1.
$$
Under suitable restrictions on the parameters $z,z',\xi$ (for instance, if
$z'=\bar z$), the above expression for $M_{z,z',\xi}(\la)$ is nonnegative for
any $\la$, so that $M_{z,z',\xi}$ is a probability measure on $\Y$. We call it
a {\it z--measure.\/} This is our main object of study. An explicit expression
for $M_{z,z',\xi}(\la)$ is given in \S1.

\subhead The measures $M^{(n)}_{z,z'}$ \endsubhead Given $n$,
restrict $M_{z,z',\xi}$ to $\Y_n\subset\Y$ and normalize it so
that the total mass of $\Y_n$ be equal to 1. Then we obtain a
probability measure on $\Y_n$ which turns out to be independent of
$\xi$; we denote this measure by $M^{(n)}_{z,z'}$. The initial
z--measure $M_{z,z',\xi}$ may be written as a mixture of the
measures $M^{(n)}_{z,z'}$ with varying $n$,
$$
M_{z,z',\xi}=\sum_{n=0}^\infty \pi(n)\,M^{(n)}_{z,z'}\,,
$$
where the coefficients
$$
\pi(n)=(1-\xi)^{zz'}\,\frac{(zz')(zz'+1)\dots(zz'+n-1)}{n!}\,\xi^n
$$
are precisely the weights of the negative binomial distribution on $\Z_+$ with
suitable parameters.

\subhead Frobenius coordinates\endsubhead We need the {\it Frobenius
notation\/} for Young diagrams:
$$
\la=(p_1,\dots,p_d\mid q_1,\dots,q_d),
$$
where $d$ is the number of diagonal boxes in $\la$, $p_i$ is the
number of boxes in the $i$th row to the right of the diagonal, and
$q_i$ is the number of boxes in the $i$th column below the
diagonal. Note that
$$
p_1>\dots>p_d\ge0, \qquad q_1>\dots>q_d\ge0.
$$
An advantage of the Frobenius notation, as compared to the
conventional notation $\la=(\la_1,\la_2,\dots)$, is its obvious
symmetry with respect to transposition of diagrams. The $p_i$'s
and $q_i$'s are called the {\it Frobenius coordinates\/} of the
Young diagram $\la$.

\subhead Asymptotic problems for random diagrams \endsubhead Given
a probability measures on Young diagrams, one may speak about {\it
random\/} Young diagrams. A problem of interest is to study the
asymptotic behavior of $M^{(n)}_{z,z'}$--random diagrams as
$n\to\infty$. In the present paper we are dealing with a different
but closely related problem: the asymptotics of
$M_{z,z',\xi}$--random diagrams as $\xi\nearrow1$ (the parameters
$z,z'$ remain fixed).

There is a number of different limit regimes of the asymptotics.
Here we discuss three of them: one for the largest Frobenius
coordinates, one for the smallest Frobenius coordinates, and one
for the Frobenius coordinates of intermediate growth.

It is an interesting question how the asymptotics of $M_{z,z',\xi}$ is related
to that of $M_{z,z'}^{(n)}$ in each of these regimes. For the first regime the
answer is known: the limiting random point processes (i.e., measures on point
configurations) are different by the multiplication by an independent random
scaling factor, see \cite{BO2, \S5} and \cite{BO1, \S6}. For the third regime,
the computation of \cite{Bor1, \S4.2-4.3}, see also \cite{BO1, \S11}, suggests
that the asymptotic behavior of the p-coordinates is the same for both
measures. However, this computation is rather involved, and it would be nice to
have a simpler argument which would also extend to the joint asymptotics of p-
and q-coordinates. No claims of this kind have been proved yet regarding the
second regime, but we believe that the corresponding asymptotics of
$M_{z,z',\xi}$ and $M_{z,z'}^{(n)}$ is also the same in this case.

\subhead Asymptotics of largest Frobenius coordinates
\cite{BO2}\endsubhead In the first limit regime, we look at the
largest Frobenius coordinates $p_1>p_2>\dots$ and $q_1>q_2>\dots$.
These are random variables depending on $\xi$ as a parameter. As
$\xi\nearrow1$, we need to normalize them, and the suitable
normalization consists in multiplying all the coordinates by
$(1-\xi)$. In the limit we obtain a couple of random infinite
sequences of decreasing real numbers, which may be also
interpreted as a random point configuration on the punctured line
$\R\setminus\{0\}$, or as a random point process on
$\R\setminus\{0\}$. This process was studied in our previous paper
\cite{BO2}. We showed that its correlation functions have
determinantal form with a kernel, which we called the {\it
Whittaker kernel,\/} because it is expressed through the classical
Whittaker function.

\subhead Limit behavior of smallest Frobenius coordinates
\endsubhead In the second limit regime, we examine the smallest Frobenius
coordinates $p_d<p_{d-1}<\dots$ and $q_d<q_{d-1}<\dots$. Again,
these are random variables depending on $\xi$, but now no
normalization is required. In the limit we obtain a couple of
random infinite increasing sequences of nonnegative integers, say
$$
0\le a_1<a_2<\dots, \qquad 0\le b_1<b_2<\dots,
$$
which can be conveniently interpreted as a random point
configuration $X$ on the lattice $\Z':=\Z+\tfrac12$ of
half--integers,
$$
X=(\dots,\; -b_2-\tfrac12,\; -b_1-\tfrac12,\; a_1+\tfrac12,\;
a_2+\tfrac12,\; \dots).
$$
Thus we get a random point process on $\Z'$, which describes the
limit behavior of the random Young diagrams near the diagonal.

A different but equivalent picture of the same limit regime is
obtained as follows. Set
$$
\unX=(X\cap\Z'_+)\cup(\Z'_-\setminus X),
$$
where
$$
\Z'_-=\{\dots,-\tfrac52,-\tfrac32,-\tfrac12\}, \qquad
\Z'_+=\{\tfrac12,\tfrac32,\tfrac52,\dots\}.
$$
That is, viewing the points of $X$ as ``particles'' and those of
$\Z'\setminus X$ as ``holes'', the configuration $\unX$ is formed
by the particles in $\Z'_+$ and the holes in $\Z'_-$.

One of the main results of the present paper is a description of
both random processes on $\Z'$. We show that the correlation
functions for each of these two processes are given in terms of a
rather simple kernel on $\Z'\times\Z'$, which is expressed through
the Euler gamma function. We call it the {\it gamma kernel\/}
(there are two versions of the kernel which correspond to the
random configurations $X$ and $\unX$, respectively). Similarly to
the Whittaker kernel, the gamma kernel depends on the parameters
$z,z'$. The version corresponding to the random configuration
$\unX$ has the form
$$
\unK^\gam(x,y\mid z,z')=\frac{\sin(\pi z)\sin(\pi
z')}{\pi\sin(\pi(z-z'))} \cdot\frac{\Cal P(x)\Cal Q(y)-\Cal
Q(x)\Cal P(y)}{x-y}
$$
where $x,y\in\Z'$ and
$$
\Cal P(x) =\frac{\Ga(z+x+\tfrac12)}
{\sqrt{\Ga(z+x+\tfrac12)\Ga(z'+x+\tfrac12)}}\,, \qquad \Cal Q(x)
=\frac{\Ga(z'+x+\tfrac12)}
{\sqrt{\Ga(z+x+\tfrac12)\Ga(z'+x+\tfrac12)}}\,.
$$
The $m$--particle correlation function ($m=1,2,\dots$) for the
random configuration $\unX$ is given by
$$
\Prob\{X\supset (x_1,\dots,x_m)\}=\det\limits_{1\le i,j\le
m}[\unK^\gam(x_i,x_j\mid z,z')].
$$
Here $(x_1,\dots,x_m)$ is an arbitrary collection of distinct
points in $\Z'$.

The correlation functions for the random configuration $X$ have
the same determinantal form, only  $\unK^\gam(x,y\mid z,z')$ is
replaced by another version of the kernel, $K^\gam(x,y\mid z,z')$,
see Theorem 3.2 below.

\subhead Asymptotics of intermediate Frobenius coordinates \endsubhead In the
third limit regime, we consider Frobenius coordinates with intermediate growth,
that is, the $p_i$'s and $q_j$'s such that
$$
0\,\ll\; p_i,\; q_j\;\ll\,\frac1{1-\xi} \qquad \text{as $\xi\nearrow1$.}
$$
We show that in a suitable scaling limit, the asymptotics of the
intermediate Frobenius coordinates is governed by a kernel on
$\R\cup\R$ (the union of two copies of the real line, one is for
p--coordinates and the other one is for q--coordinates). We call
this kernel the {\it tail kernel.\/} The tail kernel is
translationally invariant and it is a relative of the famous {\it
sine kernel.}

Notice that the tail kernel can also be obtained, via a suitable
scaling limit transition, from two opposite directions: from the
Whittaker kernel, see \cite{BO1, \S11},
 and from the gamma kernel, see \S6 below.

\subhead Remarks \endsubhead The measures $M^{(n)}_{z,z'}$ on the
finite sets $\Y_n$ first arose in Kerov--Olshanski--Vershik
\cite{KOV1}, in connection with the problem of harmonic analysis
on the infinite symmetric group. The asymptotics of the largest
Frobenius coordinates of $M^{(n)}_{z,z'}$--random Young diagrams,
as $n\to\infty$, was studied in a series of our papers, summarized
in the survey \cite{BO1}. The result provides a description of the
decomposition of the so--called generalized regular
representations of the infinite symmetric group on irreducible
components.

The z--measures $M_{z,z',\xi}$ were introduced in \cite{BO2}. They
enter a wider class of {\it Schur measures\/} introduced soon
after by Okounkov \cite{Ok2}. The z--measures initially served as
a technical tool which allowed us to rederive the main results of
\cite{Bor2}, \cite{Bor3} on the limits of the measures
$M^{(n)}_{z,z'}$ in a simpler way. However, the z--measures are
also interesting in their own right.

The idea of mixing the measures $M^{(n)}_{z,z'}$ and replacing the
large $n$ limit by the $\xi\nearrow1$ limit is similar to the idea
of passing to a grand canonical ensemble in statistical mechanics.
A parallelism between models of statistical mechanics and those of
asymptotic combinatorics was emphasized by Vershik \cite{V}.

\subhead Comparison with the Plancherel measure \endsubhead When
the parameters $z,z'$ go to infinity, the measure $M^{(n)}_{z,z'}$
degenerates to the {\it Plancherel measure\/} $M^{(n)}$ on $\Y_n$.
Similarly, when $z,z'$ go to infinity and $\xi$ goes to $+0$, in
such a way that $zz'\xi$ tends to a limit $\th>0$, the measure
$M_{z,z',\xi}$ on $\Y$ degenerates to the {\it poissonized\/}
Plancherel measure $M_\th$ with parameter $\th$. The latter
measure is a mixture of the measures $M^{(n)}$, where the mixing
distribution on the $n$'s is the Poisson distribution (the weight
of $n$ equals $e^{-\th}\th^n/n!$). The large $n$ limit of the
measures $M^{(n)}$ can be effectively replaced by the large $\th$
limit of the measures $M_\th$. Due to nice properties of the
Poisson distribution, both kinds of limit transition turn out to
be strictly equivalent in various asymptotic regimes (see
Baik-Deift-Johansson \cite{BDJ}, Borodin-Okounkov-Olshanski
\cite{BOO}, Johansson \cite{J1}).

An important difference between the Plancherel measures and the
z--measures is that the random Plancherel diagrams have a limit
form, as $n\to\infty$ or $\th\to\infty$ (see Vershik--Kerov
\cite{VK1}, \cite{VK2}, Logan-Shepp \cite{LS}), while no such form
exists for the z--measures. On the other hand, the statement of
the asymptotic problem concerning the smallest Frobenius
coordinates is the same for both kinds of measures, and the
answers are formulated in similar terms: for the Plancherel
measure, the role of the gamma kernel is played by  the {\it
discrete sine kernel\/}  with parameter 0 (see \cite{BOO},
especially Remark 1.8). Notice that the latter kernel is the
degeneration of the gamma kernel as the parameters $z,z'$ go to
infinity.

\subhead Comparison with the measures given by the Ewens
sampling formula \endsubhead The {\it Ewens sampling formula\/}
determines a one--parameter family of probability measures on
$\Y_n$ for each $n=1,2,\dots$:
$$
\ESF^{(n)}_t(\la)=\frac{t^{\ell(\la)}}{(t)_n}\,\ESF^{(n)}_1(\la),
\qquad \la\in\Y_n,
$$
where $\ell(\la)$ is the number of nonzero rows in $\la$, $t>0$
is the parameter, and
$$
\ESF^{(n)}_1(\la)=\frac{\text{the number of permutations in
$S_n$ with cycle structure $\la$}}{n!}
$$

There is a wide literature concerning these measures, see, e.g.,
the encyclopedic article Tavar\'e--Ewens \cite{TE}. As shown in
Kerov--Olshanski--Vershik \cite{KOV1}, \cite{KOV2}, both the
measures $\ESF^{(n)}_t$ and the measures $M^{(n)}_{z,z'}$ are
involved in harmonic analysis on the infinite symmetric group,
but they refer to different ``levels'', the ``group level'' and
the ``dual level'', respectively. Namely, the measures
$\ESF^{(n)}_t$  determine certain probability measures on a
compactification $\frak S$ of the infinite symmetric group,
while the measures $M^{(n)}_{z,z'}$ determine the so--called
spectral measures on the dual object to the infinite symmetric
group. The measures on $\frak S$ are used to build $L^2$ Hilbert
spaces where the so--called generalized regular representations
are realized, while the spectral measures govern the
decomposition of those representations into irreducibles.

The large $n$ limits of the measures $\ESF^{(n)}_t$ in various
regimes were extensively studied, see, e.g., the monograph by
Arratia, Barbour, and Tavar\'e \cite{ABT}. At the first glance,
the results look quite different as compared with our results
for the measures $M^{(n)}_{z,z'}$ or $M_{z,z',\xi}$.
Nevertheless, it seems to us that a detailed comparison of both
families of measures may be of interest since it could lead to a
better understanding of the nature of probabilistic models
related to partitions.

\subhead The zw--measures on signatures \endsubhead By a {\it
signature of length $N$\/}, where $N=1,2\dots$, we mean an
ordered $N$--tuple of nonincreasing integers
$$
\la=(\la_1\ge\dots\ge\la_N), \qquad \la_i\in\Z.
$$
Let $\SGN(N)$ be the set of all such $\la$'s. This is a countable
set. There is a one--to--one correspondence
$\la\leftrightarrow\chi_\la$ between signatures $\la\in\SGN(N)$
and irreducible characters of the compact group $U(N)$ of $N\times
N$ unitary matrices. The irreducible characters $\chi_\la$ are
given by the (rational) Schur functions $s_\la(u_1,\dots,u_N)$ in
the eigenvalues of a unitary matrix $U\in U(N)$. Let $H_N$ be the
Hilbert space of functions on the group $U(N)$, constant on
conjugacy classes and square integrable with respect to the
normalized Haar measure.  The characters $\chi_\la$ form an
orthonormal basis in $H_N$. Equivalently, $H_N$ can be realized as
the space of symmetric functions on the torus $\T^N$ (the product
of $N$ copies of the unit circle $\T\subset\C$), square integrable
with respect to the measure
$$
\frac1{N!}\, \prod_{1\le i<j\le N}|u_i-u_j|^2 du,
$$
where $du$ is the normalized invariant measure on the torus.

We define a family $\{f_{z,w\mid N}\}$ of vectors in $H_N$, where
$z,w$ are complex parameters, and we set
$$
M_{z,z',w,w'\mid N}(\la)=\frac{(f_{z,w\mid N},
\chi_\la)(\chi_\la,f_{\overline{w'},\, \overline{z'}\mid N})}
{(f_{z,w\mid N},f_{\overline{w'}, \,\overline{z'}\mid N})}\,,
\qquad \la\in\SGN(N),
$$
where $(\,\cdot\,,\,\cdot\,)$ is the inner product in $H_N$ and
$(z',w')$ is one more couple of complex numbers. An explicit
expression for $ M_{z,z',w,w'\mid N}(\la)$ is given in \S7. Under
suitable restrictions on the quadruple $(z,z',w,w')$, this
expression determines, for any $N$, a probability measure on
$\SGN(N)$, which we call the {\it zw-measure.\/} For instance, the
zw--measures are well defined if $z'=\bar z$, $w'=\bar w$, and
$\Re(z+w)>-\frac12$.

The zw--measures arise in the problem of harmonic analysis on the
infinite--dimensional unitary group, studied in our previous
papers \cite{Ol2}, \cite{BO4}.

\subhead Large $N$ limits of the zw--measures $M_{z,z',w,w'\mid
N}$\endsubhead Any signature $\la\in\SGN(N)$ can be viewed as a
couple $(\la^+,\la^-)$ of Young diagrams subject to the condition
$\ell(\la^+)+\ell(\la^-)\le N$, where $\ell(\la^\pm)$ stands for
the number of nonzero rows in $\la^\pm$,
$$
\la=(\la^+_1,\la^+_2,\dots,0,0,\dots,-\la^-_2,-\la^-_1).
$$
A problem of interest for the zw--measures is their limit
behavior as $N\to\infty$ with the parameters $z,z',w,w'$ being
fixed. That is, we ask about the asymptotic distribution of the
Frobenius coordinates for the random diagrams $\la^+, \la^-$.
Again, one can consider (at least) three different limit
regimes: the largest, smallest or intermediate  coordinates,
respectively.

The asymptotics of the {\it largest\/} Frobenius coordinates was
studied in \cite{Ol2}, \cite{BO4}. Introducing the scaling factor
$1/N$ for the Frobenius coordinates of $\la^+$ and $\la^-$, we
obtain in the limit 4 infinite random sequences which can be
assembled in a single random point configuration living on the
real axis with two punctures. We showed that this random point
process is governed by a kernel, which we called the {\it
continuous hypergeometric kernel\/} for it is expressed through
the Gauss hypergeometric function. This result leads to a
description of the spectral decomposition of certain unitary
representations of the infinite--dimensional unitary group.

In the present paper, we are dealing with the {\it smallest\/}
Frobenius coordinates of $\la^\pm$. That is, we study the limit
structure of the boundary of the random shape $\la^\pm$ near its
diagonal. Our result is that the limit correlation functions are
again given by the gamma kernel. The appropriate parameters are
$(-z,-z')$ for $\la^+$ and $(-w,-w')$ for $\la^-$. Thus, although
in the ``first limit regime'', the correlation kernels obtained
from the z--measures and from the zw--measures are different (the
continuous hypergeometric kernel is on the next level of
complexity as compared with the Whittaker kernel), the answer in
the ``second limit regime'' is the same.

It is worth noting that the computations leading to the gamma
kernel in the case of the zw--measures $M_{z,z',w,w'\mid N}$ are
more complex than those for the z--measures $M_{z,z',\xi}$.
Instead of the Gauss hypergeometric function ${}_2F_1$, which is
involved in the proof for the z--measures, we need to manipulate
with the higher hypergeometric series --- the series ${}_3F_2$ at
the unit argument.

As for the ``third limit regime'', which concerns {\it
intermediate\/} Frobenius coordinates, the answer is conjecturally
given by the same tail kernel as for the z--measures. We do not
prove this fact rigorously but present an argument in favor of it.

\subhead The z--measures on nonnegative signatures \endsubhead
Here we define the third family of measures, which are close
relatives of the zw--measures described above. Let $\SGN^+(N)$ be
the subset of $\SGN(N)$ formed by signatures $\la$ with
$\la_N\ge0$. We call them the {\it nonnegative signatures\/}.
Equivalently, $\SGN^+(N)$ consists of Young diagrams $\la$ with
$\ell(\la)\le N$. We fix two real parameters $a>-1$, $b>-1$. Let
$H_N$ be the Hilbert space formed by symmetric functions on the
$N$--dimensional cube $[-1,1]^N$, square integrable with respect
to the  measure
$$
\prod_{1\le i<j\le N}(x_i-x_j)^2\cdot \prod_{i=1}^N
(1-x_i)^a(1+x_i)^b \cdot dx_1\dots dx_n\,, \qquad x_i\in[-1,1].
$$

In $H_N$, we consider the orthonormal basis $\{\chi^{a,b}_\la\}$
formed by the (suitably normalized) multivariate Jacobi
polynomials. Here the subscript $\la$ ranges over $\SGN^+(N)$. On
the other hand, we introduce a family $\{f_{z\mid N}\}$ of
symmetric functions on the cube, depending on a complex number
$z$, and we set
$$
M_{z,z',a,b\mid N}(\la)=\frac{(f_{z\mid N},
\chi^{a,b}_\la)(\chi^{a,b}_\la,f_{\overline{z'}\,\mid N})}
{(f_{z\mid N},f_{\overline{z'}\,\mid N})}\,,  \qquad
\la\in\SGN^+(N),
$$
where $z'$ is one more complex parameter. If $z, z'$ satisfy
certain restrictions, this gives us a probability measure on
$\SGN^+(N)$, which we call the {\it z-measure on nonnegative
signatures\/}. An explicit expression is given in \S8.

This construction is again motivated by representation theory.
Specifically, for a few special values of $(a,b)$, the (suitably
renormalized) multivariate Jacobi polynomials are the irreducible
characters of the symplectic or orthogonal groups, or else the
spherical functions on the complex Grassmannians. Then the
z--measures $M_{z,z', a,b\mid N}$ naturally emerge in the problem
of harmonic analysis for infinite--dimen\-sional analogs of these
classical groups or for the Grassmannians. \footnote{This problem
can be stated by analogy with the construction of \cite{Ol2}.
Notice that the z--measures $M_{z,z', a,b\mid N}$ with $a=b=0$ and
real $z=z'$ appeared for the first time in \cite{Pic}.}

For general $(a,b)$, there is no such direct
representation--theoretic interpretation. Nevertheless, according
to the philosophy of the modern theory of multivariate special
functions (see, e.g. Heckman's part of the book \cite{HS}), there
are good reasons to work with general parameters $(a,b)$ as well.

\subhead Large $N$ limit of the z--measures $M_{z,z',a,b\mid N}$
\endsubhead As in the case of the zw--measures, we focus on the
``second limit regime'', which, in the present case, concerns the
smallest Frobenius coordinates of the random diagrams
$\la\in\SGN^+(N)$. And once again, it turns out that the limit
point process is determined by the gamma kernel. The proof
involves rather tedious computations with the hypergeometric
series ${}_4F_3$ at the unit argument.

\subhead Asymptotics of discrete orthogonal polynomials
\endsubhead
The heart of the argument in the cases of the zw-measures
$M_{z,z',w,w'\mid N}$ and the z-measures $M_{z,z',a,b\mid N}$ is
a computation of the asymptotics of certain discrete orthogonal
polynomials of degree $N-1$ and $N$ as $N\to\infty$. Those are
the Askey--Lesky polynomials (which generalize the classical
Hahn polynomials) in the first case and the Wilson-Neretin
polynomials (which generalize the classical Racah polynomials)
in the second case. Even though the weight function in both
cases depends on four independent parameters (except for $N$),
the limits of the Christoffel--Darboux kernels in both cases are
the same (the gamma kernel), and the result depends only on two
of the four initial parameters. This suggests that the gamma
kernel (and, hence, the tail kernel which is equal to its
scaling limit) may play a universal role in asymptotics of
general discrete orthogonal polynomials.

Recall that, as it was recently shown by
Baik--Kriecherbauer--McLaughlin--Miller \cite{BKMM, \S3.1.1}, the
discrete sine kernel is the universal microscopic limit of the
Christoffel--Darboux kernels associated with generic discrete
orthogonal polynomials, near a point where the macroscopic density
function is continuous and takes any value strictly between 0 and
1.\footnote{This means that the point configurations of the
corresponding orthogonal polynomial ensemble are neither empty
(density 0) nor fully packed (density 1) near this point.}

It looks very plausible to us that the gamma kernel and the tail
kernel are universal microscopic limits, in two different
asymptotic regimes, of the Christoffel--Darboux kernels for
generic discrete orthogonal polynomials near a point where the
macroscopic density function is discontinuous, takes value 0 on
one side of this point, and takes value 1 on the other side of
this point. The two special cases considered in \S7 and \S8 below
provide some evidence in support of this conjecture.

\head 1. Definition of the z--measures \endhead

As in Macdonald \cite{Ma} we identify partitions and Young
diagrams. By $\Y_n$ we denote the set of partitions of a natural
number $n$, or equivalently, the set of Young diagrams with $n$
boxes. By $\Y$ we denote the set of all Young diagrams, that is,
the disjoint union of the finite sets $\Y_n$, where
$n=0,1,2,\dots$ (by convention, $\Y_0$ consists of a single
element, the empty diagram $\varnothing$).

Given $\la\in\Y$, let $|\la|$ denote the number of boxes of $\la$ (so that
$\la\in\Y_{|\la|}$), let $\ell(\la)$ be the number of nonzero rows in $\la$,
and let $\la'$ be the transposed diagram. For $z\in\C$, let
$$
(z)_\la=\prod_{i=1}^{\ell(\la)}(z-i+1)_{\la_i}
$$
where $(x)_k=x(x+1)\dots(x+k-1)=\Ga(x+k)/\Ga(x)$ is the
Pochhammer symbol. Note that
$$
(z)_\la=\prod_{(i,j)\in\la}(z+j-i)
$$
(product over the boxes of $\la$), which implies at once the symmetry relation
$$
(z)_\la=(-1)^{|\la|}(-z)_{\la'}.
$$

Given $\la\in\Y$, $\la\ne\varnothing$, we denote by $\chi_\la$ the
irreducible character of the symmetric group $S_{|\la|}$, indexed
by $\la$. For $n=1,2,\dots$, let $H_n$ be the space of complex
functions on $S_n$, constant on conjugacy classes. We introduce an
inner product in $H_n$ by the formula
$$
(f,g)_n=\frac1{n!}\,\sum_{s\in S_n}f(s)\overline{g(s)}.
$$
The characters $\chi_\la$ with $\la\in\Y_n$ form an orthonormal basis in $H_n$,
so that we may write
$$
H_n=\bigoplus_{\la\in\Y_n}\C\chi_\la\,, \qquad n=1,2,\dots\,.
$$
We also agree that $H_0$ is a one--dimensional vector space with basis element
denoted as $\chi_\varnothing$, $(\chi_\varnothing,\chi_\varnothing)_0=1$.

Given $z\in\C$, define a function $f^{(n)}_z\in H_n$ as follows
$$
f^{(n)}_z(s)=z^{\text{the number of cycles in $s$}}, \qquad s\in S_n, \quad
n=1,2,\dots
$$

\proclaim{Proposition 1.1 \cite{KOV2, Lemma 4.1.2}} The expansion of
$f^{(n)}_z$ in the basis $\{\chi_\la\}$ has the form
$$
f^{(n)}_z=\sum_{\la\in\Y_n}(z)_\la \frac{\dim\la}{n!}\,\chi_\la
$$
where
$$
\dim\la=\chi_\la(e).
$$
\endproclaim

For the reader's convenience we outline the proof.

\demo{Proof} Let $\La$ be the graded algebra of symmetric
functions in countably many variables, say $y_1,y_2,\dots$, and
let $\La^n$ denotes the $n$th homogeneous component of $\La$
(\cite{Ma, \S I.2}). Endow $\La$ with the canonical inner product
(\cite{Ma, \S I.4}). Consider the characteristic map
$\operatorname{ch}_n$, which is a linear isometry between $H_n$
and $\La^n$ transforming the characters $\chi_\la$ into the Schur
functions $s_\la$ (\cite{Ma, \S I.7}). First, we check that
$$
\operatorname{ch}_n(f^{(n)}_z) =\text{\;the $n$th homogeneous component of
\;$\prod_{i=1}^\infty(1-y_i)^{-z}$}.
$$
This reduces the claim of the proposition to the expansion
$$
\prod_{i=1}^\infty(1-y_i)^{-z}=\sum_{\la\in\Y}(z)_\la
\frac{\dim\la}{|\la|!}\,s_\la,
$$
which in turn can be deduced from \cite{Ma, chapter I, (4.3)}. \qed
\enddemo

As is well known, $\dim\la$ coincides with the number of standard
tableaux of shape $\la\in\Y$ (see \cite{Ma, Example I.7.3}). A
number of different explicit expressions are known for this
quantity. For instance, for any natural $k\ge\ell(\la)$,
$$
\frac{\dim\la}{|\la|!}=\det\limits_{1\le i,j\le
k}\left[\frac1{\Gamma(\la_i-i+j+1)}\right]=\frac{\prod_{1\le i<j\le
k}(\la_i-\la_j+j-i)}{\prod_{1\le i\le k}(\la_i+k-i)!}
$$
(\cite{Ma, Example I.7.6}). These formulas do not demonstrate the
symmetry $\dim\la=\dim\la'$. There are two other formulas which
are symmetric: the hook formula (\cite{Ma, Example I.5.2}) and the
expression in terms of Frobenius coordinates, see the beginning of
\S3 below.

Let us agree that
$$
f^{(0)}_z=\chi_\varnothing\in H_0\,.
$$

\proclaim{Proposition 1.2} For any $z\in\C$ and any $\xi\in(0,1)$, we have
$$
\sum_{n=0}^\infty\Vert f^{(n)}_z\Vert_n^2\, \frac{\xi^n}{n!} <+\infty
$$
so that the formal sum
$$
f_{z,\xi}:=\sum_{n=0}^\infty f^{(n)}_z\sqrt{\frac{\xi^n}{n!}}
$$
is a well--defined element of the Hilbert space
$$
H:=H_0\oplus H_1 \oplus H_2\oplus\dots
$$
\endproclaim

\demo{Proof} We will prove that
$$
\Vert f^{(n)}_z\Vert_n^2=(z\bar z)_n=(z\bar z)(z\bar z+1)\dots(z\bar z+n-1).
$$
Since the series
$$
\sum_{n=0}^\infty\frac{(z\bar z)_n}{n!}\xi^n
$$
converges for $0<\xi<1$, the claim of the proposition will readily follow.

By the definition of $f^{(n)}_z$
$$
\Vert f^{(n)}_z\Vert_n^2=(f^{(n)}_{z\bar z}, \chi_{(n)})_n
$$
where $(n)$ is the partition $(n,0,0,\dots)$ (the corresponding character is
simply the constant function 1). Then the result follows from Proposition 1.1.
\qed
\enddemo

\proclaim{Proposition 1.3} For any $z,z'\in\C$ and any $\xi\in(0,1)$, we have
$$
(f_{z,\xi},f_{\overline{z'},\xi})=(1-\xi)^{-zz'}\,,
$$
where $(\,\cdot\,,\,\cdot\,)$ is the inner product in $H$.
\endproclaim

\demo{Proof} The same argument as in the proof of Proposition 1.2 gives
$$
(f^{(n)}_{z},f^{(n)}_{\overline{z'}})_n=(zz')_n.
$$
Therefore,
$$
(f_{z,\xi},f_{\overline{z'},\xi}) =\sum_{n=0}^\infty
\left(f^{(n)}_{z},\,f^{(n)}_{\overline{z'}}\right)_n
\frac{\xi^n}{n!} =\sum_{n=0}^\infty
\frac{(zz')_n\,\xi^n}{n!}=(1-\xi)^{-zz'}\,.\qed
$$
\enddemo

\example{Definition 1.4} (i) Let $z,z'\in\C$ and $\xi\in(0,1)$. For any
$\la\in\Y$ we set
$$
M_{z,z',\xi}(\la)=\frac{(f_{z,\xi}\,,\chi_\la)\,(\chi_\la\,,
f_{\overline{z'},\xi})}{(f_{z,\xi}\,, f_{\overline{z'},\xi})}
$$
Notice that the denominator is nonzero (Proposition 1.3), so that the whole
expression makes sense. Since $\{\chi_\la\}_{\la\in\Y}$ is an orthonormal basis
in $H$, we have
$$
\sum_{\la\in\Y}M_{z,z',\xi}(\la)=1.
$$
{}From Propositions 1.2 and 1.3 we obtain an explicit expression
for $M_{z,z',\xi}$:
$$
M_{z,z'\xi}(\la) =(1-\xi)^{zz'}\,
(z)_\la(z')_\la\,\left(\frac{\dim\la}{|\la|!}\right)^2\,\xi^{|\la|}\,.
$$

(ii) Under suitable restrictions on the triple $(z,z',\xi)$ the
quantities $M_{z,z',\xi}(\la)$ are nonnegative for all $\la$. Then
$M_{z,z',\xi}$ is a probability measure on the countable set $\Y$,
which we call the {\it z--measure\/} on $\Y$ with parameters
$z,z',\xi$. The nonnegativity property holds, for instance, if
$z'=\bar z$; other sufficient conditions are given in Corollary
1.9 below.  The definition of the z--measures $M_{z,z',\xi}$ was
given in Borodin--Olshanski \cite{BO2}; see also \cite{BO3}. It is
a modification of a construction due to Kerov--Olshanski--Vershik
\cite{KOV1}, \cite{KOV2}. They z--measures enter a larger class of
{\it Schur measures\/} as defined by Okounkov \cite{Ok2}.

\endexample

\example{Example 1.5} Assume $z=k$, $z'=l$, where $k,l$ are natural numbers.
Then $(z)_\la$ vanishes unless $\ell(\la)\le k$, and likewise $(z')_\la$
vanishes unless $\ell(\la)\le l$. If $\ell(\la)\le\min(k,l)$ then both
$(z)_\la$ and $(z')_\la$ are strictly positive. It follows that $M_{k,l,\xi}$
is a measure supported by diagrams $\la$ with at most $\min(k,l)$ rows. This
z--measure can be obtained by the following construction.

Let $S(\C^k\otimes\C^l)$ be the symmetric algebra of the vector
space $\C^k\otimes\C^l$. This is a graded space. Let $A_\xi$ be
the operator in $S(\C^k\otimes\C^l)$ taking value $\xi^n$ on the
$n$th homogeneous component. On the other hand, as a bi--module
over $GL(k,\C)\times GL(l,\C)$, the space $S(\C^k\otimes\C^l)$
is the multiplicity free direct sum of irreducible bi--modules
of the form $V_{\la,k}\otimes V_{\la,l}$, where $\la$ ranges
over the set of Young diagrams with $\ell(\la)\le\min(k,l)$ and
$V_{\la,k}$ denotes the irreducible polynomial
$GL(k,\C)$--module indexed by $\la$. Given $\la$ with
$\ell(\la)\le\min(k,l)$, denote by $I_\la$ the projection onto
the component $V_{\la,k}\otimes V_{\la,l}$. Then we have
$$
M_{k,l,\xi}(\la)=\frac{\tr(A_\xi I_\la)}{\tr A_\xi}\,, \qquad
\ell(\la)\le\min(k,l).
$$

A closely related interpretation is as follows. Consider the set
$\Mat(k,l;\Z_+)$ of $k\times l$ matrices with entries in $\Z_+$.
The Robinson--Schensted--Knuth algorithm (RSK, for short)
determines a projection of $\Mat(k,l;\Z_+)$ onto the set of
Young diagram with at most $\min(k,l)$ rows (see e.g. Sagan
\cite{Sa, Theorem 4.8.2}). Let $\wt M_{k,l,\xi}$ be the
probability measure on $\Mat(k,l;\Z_+)$ defined by the condition
that the matrix entries are independent random variables
distributed according to the geometric distribution with
parameter $\xi$. Then the push--forward of $\wt M_{k,l,\xi}$
under RSK is $M_{k,l,\xi}$.

Asymptotics of the first part of random partitions distributed
according to $M_{k,l,\xi}$, has been thoroughly studied by
Johansson \cite{J1}.

\endexample

\example{Example 1.6} Once again, let $k,l$ be two natural
numbers, and take $z=k$, $z'=-l$. Then $(z)_\la(z')_\la$ vanishes
unless $\ell(\la)\le k$ and $\ell(\la')\le l$, that is, $\la$ must
be contained in the rectangular shape of size $k\times l$. If this
condition is satisfied then the sign of $(z)_\la(z')_\la$ equals
$(-1)^{|\la|}$. Assume now that $\xi<0$ (we temporarily abandon
the restriction $\xi\in(0,1)$). Then the factor $\xi^{|\la|}$ in
Definition 1.4 will compensate the oscillation of the sign of
$(z)_\la(z')_\la$, and we again obtain a probability measure,
$M_{k,-l,\xi}$. Note that it is supported by a finite set of Young
diagrams.

Both interpretations of the measure $M_{k,l,\xi}$ given in
Example 1.5 can be extended to the measure $M_{k,-l,\xi}$, with
suitable modifications. Namely, the symmetric algebra $S(\C^
k\otimes\C^l)$ is replaced by the exterior algebra $\wedge(\C^
k\otimes\C^l)$. Let $A'_\xi$ be the operator in this graded
space taking value $(-\xi)^n$ on the $n$th homogeneous
component. The exterior algebra decomposes into irreducible
bi--modules of the form $V_{\la,k}\otimes V_{\la',l}$. Let
$I_\la$ denote the projection onto $V_{\la,k}\otimes
V_{\la',l}$. Then
$$
M_{k,-l,\xi}(\la)=\frac{\tr(A'_\xi I_\la)}{\tr A'_\xi}\,, \qquad \ell(\la)\le
k, \quad \ell(\la')\le l.
$$

Further, consider the (finite) set $\Mat(k,l;0/1)$ of $k\times l$
matrices with entries in $\{0,1\}$. We equip $\Mat(k,l;0/1)$ with
the probability measure $\wt M_{k,-l,\xi}$ such that the matrix
entries are independent and identically distributed according to
$$
\Prob(0)=\frac{1}{1+|\xi|}, \quad \Prob(1)=\frac{|\xi|}{1+|\xi|}.
$$
Instead of the Robinson--Schensted--Knuth algorithm we apply its
dual version (dRSK), see \cite{Sa, Theorem 4.8.5}. Taking the
push--forward of $\wt M_{k,-l,\xi}$ with respect to dRSK we
obtain $M_{k,-l,\xi}$.

Asymptotics of the first part of random partitions distributed
according to $M_{k,-l,\xi}$, has been thoroughly studied by
Gravner--Tracy--Widom \cite{GTW}.
\endexample

\example{Example 1.7} Let the parameters $z,z',\xi$ vary in such a way that
$$
|z|\to\infty, \quad |z'|\to\infty, \quad \xi\to0, \quad zz'\xi\to\theta>0.
$$
Then we obtain in the limit the {\it poissonized Plancherel measure\/} with
parameter $\theta$,
$$
M_\theta(\la)=e^{-\theta}\,\left(\frac{\dim\la}{|\la|!}\right)^2\theta^{|\la|}.
$$
Asymptotics of the Plancherel measure has been studied by many
authors, see e.g. \cite{LS}, \cite{VK1}, \cite{VK2}, \cite{BDJ},
\cite{BOO}, \cite{J2}, \cite{Ok3} and references therein.
\endexample

\proclaim{Proposition 1.8} Let $z,z'$ be nonzero complex numbers.
The quantity $(z)_\la(z')_\la$ is nonnegative for any $\la\in\Y$
if and only if one of the following three conditions holds:

{\rm(i)} The numbers $z,z'$ are not real and are conjugate to each
other.

{\rm(ii)} Both $z,z'$ are real and are contained in the same open
interval of the form $(m,m+1)$, where $m\in\Z$.

{\rm(iii)} One of the numbers $z,z'$ {\rm(}say, $z${\rm)} is a
nonzero integer while $z'$ has the same sign and, moreover,
$|z'|>|z|-1$.
\endproclaim

\demo{Proof} Consider two cases: (1) both $z,z'$ are not integers;
(2) at least one of $z,z'$ is an integer.

(1) In this case, the quantity $(z)_\la(z')_\la$ does not vanish.
It is strictly positive for all $\la\in\Y$ if and only if
$(z+k)(z'+k)>0$ for any integer $k$, which is equivalent to
$(z,z')$ satisfying (i) or (ii).

(2) Without loss of generality we may assume that either $z$ is an
integer and $z'$ is not, or both $z,z'$ are integers and $|z|\le
|z'|$. Next, by virtue of the symmetry
$(z)_\la(z')_\la=(-z)_{\la'}(-z')_{\la'}$, we may assume
$z=m=1,2,\dots$ (note that $z=0$ is excluded by the hypothesis).
Then $(z)_\la$ vanishes if $\ell(\la)>m$, and is strictly positive
if $\ell(\la)\le m$. Therefore, the quantity $(z)_\la(z')_\la$ is
nonnegative for all $\la$ if and only if $(z')_\la\ge0$ for all
$\la$ with $\ell(\la)\le m$, which means that $z'$ must be a real
number $> m-1$. (Note that $z'\ne m-1$ because of the assumption
$|z|\le |z'|$.) \qed
\enddemo

\proclaim{Corollary 1.9} Let $z,z'$ satisfy one of the
conditions (i), (ii), (iii) of Proposition 1.8, and let
$0<\xi<1$. Then the z--measure $M_{z,z',\xi}$ with parameters
$z,z',\xi$ is well defined as a probability measure.
\endproclaim

Notice the symmetry relation
$$
M_{z,z',\xi}(\la)=M_{-z,-z',\xi}(\la').
$$

Henceforth we assume that the parameter $\xi$ belongs to the open interval
$(0,1)$.

\head 2. The hypergeometric and gamma kernels (first form) \endhead

Let $\Z'$ denote the lattice of proper half--integers:
$$
\Z'=\Z+\tfrac12=\{\pm\tfrac12, \pm\tfrac32, \pm\tfrac52, \dots\}.
$$
Consider the space of all subsets $X\subset\Z'$. By assigning to
any $X\subset\Z'$ its characteristic function we identify that
space with the space $\{0,1\}^{\Z'}$ of all doubly infinite binary
sequences indexed by elements of the lattice $\Z'$:
$$
(\dots, a_{-3/2}, a_{-1/2}\mid a_{1/2}, a_{3/2},\dots), \qquad a_x\in\{0,1\}
\quad \forall x\in\Z'.
$$
We endow the space $\{0,1\}^{\Z'}$ with the product topology, which makes it a
compact topological space.

To any diagram $\la\in\Y$ we assign a subset $\unX(\la)\subset\Z'$,
$$
\unX(\la)=\{\la_i-i+\tfrac12\mid i=1,2,\dots\},
$$
which we identify with the corresponding binary sequence
$(a\,_x(\la))_{x\in\Z'}$ (so that $a_x=1$ if $x$ equals
$\la_i-i+\tfrac12$ for some $i$, and $a_x=0$ otherwise). Thus, we
obtain an embedding $Y\hookrightarrow \{0,1\}^{\Z'}$. For
instance, the empty diagram turns into the binary sequence
$(\dots111\mid000\dots)$, and the diagram $\la=(3,1)\in\Y_3$ turns
into the binary sequence $(\dots11101\mid00100\dots)$. The binary
sequence $(a_x(\la))_{x\in\Z'}$  has a simple geometric meaning:
given $k=1,2,\dots$, the digit $a_{\pm(k-1/2)}$ is 1 or 0
depending on whether the $k$th segment of the boundary of $\la$
above/below the diagonal is vertical or horizontal.

Note that image of $\Y$ is dense in $\{0,1\}^{\Z'}$, so that
$\{0,1\}^{\Z'}$ is  a compactification of the discrete space $\Y$.

\example{Definition 2.1} Let $P$ be an arbitrary probability
measure on the compact space $\{0,1\}^{\Z'}$. The {\it
$m$--point correlation function\/} ($m=1,2,\dots$) of $P$,
denoted as $\rho_m(\,\cdot\mid P)$, is defined on $m$--point
subsets $X=\{x_1,\dots,x_m\}\subset\Z'$. The value
$\rho_m(x_1,\dots,x_m\mid P)$ at $X$ is the probability that the
random (with respect to $P$) set contains $X$. Equivalently,
this is the probability that the random (with respect to $P$)
binary sequence has 1's at the positions $x_1,\dots,x_m$. \qed
\endexample

Notice that $P$ is uniquely determined by its correlation
functions. Indeed, using the inclusion/exclusion principle we can
compute the $P$--measure of any cylinder set of the form
$\{(a_x)\mid a_{y_1}=\epsi_1,\dots,a_{y_m}=\epsi_m\}$ with
arbitrary $y_1,\dots,y_m\in\Z'$ and $\epsi_1,\dots,\epsi_m=0,1$.

It turns out that the correlation functions of the z--measures can
be explicitly computed.

\proclaim{Theorem 2.2} Assume that both $z,z'$ are not integers.
That is, $(z,z')$ is subject to one of the conditions {\rm(i)},
{\rm(ii)} of Proposition 1.8, but not to the condition {\rm(iii)}.
Let $\unP_{z,z',\xi}$ be the push--forward of the z--measure
$M_{z,z',\xi}$ under the embedding $\la\mapsto\unX(\la)$ of the
discrete space $\Y$ into the compact space $\{0,1\}^{\Z'}$.

The correlation functions of $\unP_{z,z',\xi}$, as defined in Definition 2.1,
have determinantal form
$$
\gather
\rho_m(x_1,\dots,x_m\mid \unP_{z,z',\xi})=\det_{1\le i,j\le
m}[\unK(x_i,x_j\mid z,z',\xi)], \\
m=1,2,\dots, \quad x_1,\dots,x_m\in\Z',
\endgather
$$
where $\unK(x,y\mid z,z',\xi)$ is a function on $\Z'\times\Z'$ not depending on
$m$. Specifically,
$$
\unK(x,y\mid z,z',\xi)=\frac{P(x)Q(y)-Q(x)P(y)}{x-y}\,,
$$
with
$$
\multline P(x)=P(x\mid z,z',\xi)=(zz')^{1/4}\xi^{x/2}(1-\xi)^{(z+z')/2}\\
\times\left(\frac{\Ga(z+x+\frac12)\Ga(z'+x+\frac12)}
{\Ga(z+1)\Ga(z'+1)}\right)^{1/2}
\cdot\frac{F(-z,-z';x+\frac12;\tfrac{\xi}{\xi-1})}{\Ga(x+\frac12)}
\endmultline
$$
$$
\multline
Q(x)=Q(x\mid z,z',\xi)=(zz')^{3/4}\xi^{(x+1)/2}(1-\xi)^{(z+z')/2-1}\\
\times\left(\frac{\Ga(z+x+\frac12)\Ga(z'+x+\frac12)}{\Ga(z+1)\Ga(z'+1)}\right)^{
1/2}
\cdot\frac{F(-z+1,-z'+1;x+\frac32;\tfrac{\xi}{\xi-1})}{\Ga(x+\frac32)}\,,
\endmultline
$$
where $F(a,b;c;w)$ stands for the Gauss hypergeometric function.
\endproclaim

\demo{Comments} 1. The ratio $F(a,b;c;w)/\Ga(c)$ is an entire
function in the parameter $c$, see Erdelyi \cite{Er1, 2.1.6}.
Next, under our assumptions on the parameters $z,z'$,
$$
\frac{\Ga(z+x+\frac12)\Ga(z'+x+\frac12)}{\Ga(z+1)\Ga(z'+1)}>0.
$$
This implies that $P(x)$ and $Q(x)$ are well defined on the whole lattice
$\Z'$.

2. Moreover, the expressions of the functions $P(x)$, $Q(x)$ are also well
defined in a neighborhood of $\Z'$ in $\C$, and these are analytic functions.
This makes it possible to define the value of ratio $(P(x)Q(y)-Q(x)P(y))/(x-y)$
on the diagonal $x=y$, by making use of the l'Hospital rule.

3. Notice that
$$
Q(x\mid z,z',\xi)=\left(\frac{zz'}{(z-1)(z'-1)}\right)^{1/4}\, P(x+1\mid z-1,
z'-1,\xi).
$$

4. We call $\unK(x,y\mid z,z',\xi)$ the {\it discrete hypergeometric kernel.}
\enddemo

\demo{Proof of Theorem 2.2}  As is shown below (Corollary 4.3),
Theorem 2.2 is equivalent to Theorem 3.2, and the latter theorem
was proved in Borodin--Olshanski \cite{BO2}. On the other hand,
Theorem 2.2 can be proved directly, see Okounkov \cite{Ok1} and
Borodin-Okounkov \cite{BOk, Example 3}.\qed
\enddemo

Recall that the parameter $\xi$ of the z--measure ranges over
the open interval $(0,1)$. What happens when $\xi$ tends to one
of the endpoints 0, 1? {}From the definition of the z--measures
it easily follows that as $\xi$ tends to 0, the z--measure tends
to the Dirac measure at $\varnothing$, while the limit as $\xi$
tends to 1 is the zero measure:
$$
\lim_{\xi\nearrow1}M_{z,z',\xi}(\la)=0, \qquad \forall \la\in\Y.
$$
However, the $\xi\nearrow1$ limit becomes nontrivial when
instead of $\Y$ we take its compactification
$\{0,1\}^{\Z'}\supset\Y$.

\proclaim{Theorem 2.3} Let $(z,z')$ and $\unP_{z,z',\xi}$ be as
in Theorem 2.2. As $\xi\nearrow1$, the measures
$\unP_{z,z',\xi}$ weakly converge to a probability measure
$\unP_{z,z'}^\gam$ on $\{0,1\}^{\Z'}$. The correlation functions
of the limit measure have determinantal form,
$$
\gather \rho_m(x_1,\dots,x_m\mid \unP_{z,z'}^\gam)=\det_{1\le i,j\le
m}[\unK^\gam(x_i,x_j\mid z,z')], \\
m=1,2,\dots, \quad x_1,\dots,x_m\in\Z',
\endgather
$$
where $\unK^\gam(x,y\mid z,z',\xi)$ is a function on $\Z'\times\Z'$ not
depending on $m$. Specifically,
$$
\multline \unK^\gam(x,y\mid z,z') =\frac{\sin(\pi z)\sin(\pi
z')}{\pi\sin(\pi(z-z'))} \\
\times \left\{\Ga(z+x+\tfrac12)\Ga(z'+x+\tfrac12)
\Ga(z+y+\tfrac12)\Ga(z'+y+\tfrac12)\right\}^{-1/2}\\
\times
\frac{\Ga(z+x+\tfrac12)\Ga(z'+y+\tfrac12)-
\Ga(z'+x+\tfrac12)\Ga(z+y+\tfrac12)}{x-y}
\endmultline
$$
\endproclaim

\demo{Comments} 1. The expression in the curved brackets is
strictly positive because of our assumptions on the parameters
$z,z'$.

2. If $z=z'$, which is only possible when $z=z'\in\R\setminus\Z$, then the
above expression takes a simpler  form
$$
\unK^\ps(x,y\mid z)=\left(\frac{\sin(\pi z)}{\pi}\right)^2\,
\frac{\psi(z+x+\tfrac12)-\psi(z+y+\tfrac12)}{x-y}\,,
$$
where $\psi(u)=\Gamma'(u)/\Gamma(u)$ is the logarithmic derivative of the
$\Gamma$--function.

3. On the diagonal $x=y$ we have
$$
\gather
\unK^\gam(x,x\mid z,z')=\frac{\sin(\pi z)\sin(\pi z')}
{\pi\sin(\pi(z-z'))}\, (\psi(z+x+\tfrac12)-\psi(z'+x+\tfrac12))\\
\unK^\ps(x,y\mid z)\bigg|_{x=y}=\left(\frac{\sin(\pi z)}{\pi}\right)^2\,
\psi'(z+x+\tfrac12).
\endgather
$$

4. We call $\unK^\gam(x,y\mid z,z')$ and $\unK^\ps(x,y\mid z)$ the
{\it gamma kernel\/} and the {\it psi kernel\/}, respectively.

\enddemo

\demo{Proof of Theorem 2.3} We will show that the discrete hypergeometric
kernel $\unK(x,y\mid z,z',\xi)$ of Theorem 2.2 has a pointwise limit as
$\xi\nearrow 1$, and the result is the gamma kernel. This will imply Theorem
2.3. (Notice, however, that the functions $P(x\mid z,z',\xi)$ and $Q(x\mid
z,z',\xi)$, in general, do not have limits as $\xi\nearrow1$.)

We use the formula (see Erdelyi \cite{Er1, 2.1.4 (17)})
$$
\gather \frac1{\Ga(c)}\, F(a,b;c;w)= \frac{\Ga(b-a)(-w)^{-a}}{\Ga(b)\Ga(c-a)}\,
F(a,1-c+a;1-b+a;w^{-1})\\
+ \frac{\Ga(a-b)(-w)^{-b}}{\Ga(a)\Ga(c-b)}\, F(b,1-c+b;1-a+b;w^{-1}), \qquad
w\in\C\setminus[0,+\infty)
\endgather
$$
For fixed $a,b,c$ and large negative $w$, we write
$$
F(a,1-c+a;1-b+a;w^{-1})=1+O(w^{-1}), \quad
F(b,1-c+b;1-a+b;w^{-1})=1+O(w^{-1}),
$$
which gives
$$
\frac1{\Ga(c)}\, F(a,b;c;w)=
\frac{\Ga(b-a)(-w)^{-a}}{\Ga(b)\Ga(c-a)}\, (1+O(w^{-1}))+
\frac{\Ga(a-b)(-w)^{-b}}{\Ga(a)\Ga(c-b)}\, (1+O(w^{-1})), \tag
2.1
$$
Specializing this simple estimate to
$$
a=-z, \qquad b=-z', \qquad c=x+\frac12, \qquad w=\frac{\xi}{\xi-1}
$$
and to
$$
a=-z+1, \qquad b=-z'+1, \qquad c=x+\frac32, \qquad w=\frac{\xi}{\xi-1}
$$
we obtain (below we denote by $\de_1,\de_2,\dots$ suitable quantities of the
type $1+O(1-\xi)$ whose precise form is unessential)
$$
\gather P(x\mid
z,z',\xi)=(zz')^{1/4}\left(\frac{\Ga(z+x+\tfrac12)\Ga(z'+x+\tfrac12)}
{\Ga(z+1)\Ga(z'+1)}\right)^{1/2}\\
\times\left\{
\frac{\Ga(z-z')(1-\xi)^{(z'-z)/2}}{\Ga(-z')\Ga(z+x+\tfrac12)}\,\de_1 +
\frac{\Ga(z'-z)(1-\xi)^{(z-z')/2}}{\Ga(-z)\Ga(z'+x+\tfrac12)}\,\de_2
\right\}
\\
Q(y\mid z,z',\xi) =(zz')^{3/4}\left(\frac{\Ga(z+y+\tfrac12)\Ga(z'+y+\tfrac12)}
{\Ga(z+1)\Ga(z'+1)}\right)^{1/2}\\
\times\left\{
\frac{\Ga(z-z')(1-\xi)^{(z'-z)/2}}{\Ga(-z'+1)\Ga(z+y+\tfrac12)}\,\de_3 +
\frac{\Ga(z'-z)(1-\xi)^{(z-z')/2}}{\Ga(-z+1)\Ga(z'+y+\tfrac12)}\,\de_4 \right\}
\endgather
$$
Substituting  these expressions into $P(x)Q(y)$ one sees that the
term involving the factor $(1-\xi)^{z-z'}$ or $(1-\xi)^{z'-z}$
will cancel with the corresponding term in $Q(x)P(y)$, within a
quantity of the form $(1-\xi)^{\pm(z-z')}O(1-\xi)$. Such a
quantity is negligible, because $|\Re(z-z')|<1$, as it follows
from our assumptions on $z,z'$. Thus, only terms not involving the
factors $(1-\xi)^{\pm(z-z')}$ survive in $P(x)Q(y)-Q(x)P(y)$.
Writing these terms down we get
$$
\gather P(x\mid z,z',\xi)\,Q(y\mid z,z',\xi)-Q(x\mid z,z',\xi)\,P(y\mid
z,z',\xi)\\
=\frac{\sin(\pi z)\sin(\pi z')}{\pi\sin(\pi(z-z'))}
\,\left\{\Ga(z+x+\tfrac12)\Ga(z'+x+\tfrac12)
\Ga(z+y+\tfrac12)\Ga(z'+y+\tfrac12)\right\}^{-1/2}\\
\times
\left\{\Ga(z+x+\tfrac12)\Ga(z'+y+\tfrac12)
-\Ga(z'+x+\tfrac12)\Ga(z+y+\tfrac12)\,+\,o(1)\right\}.
\endgather
$$
This proves the claim of the theorem for $x\ne y$ and $z\ne z'$.
To remove the restriction $x\ne y$ we remark that the above
formula holds not only for $x,y$ on the lattice $\Z'$ but also in
a suitable neighborhood $U$ of the lattice $\Z'$ in $\C$.
Moreover, one can prove that the remaining term $o(1)$ admits a
uniform bound provided that $x,y$ range over compact subsets in
$U\times U$. Thus, as $\xi\to1$, the left--hand side (which is a
holomorphic function in $U\times U$, vanishing on the diagonal
$x=y$) converges to the right--hand side with the remaining term
(which has the same vanishing property) removed, uniformly on
compact sets. This makes it possible to remove the indeterminacy
on the diagonal $x=y$ using the L'Hospital rule.

Finally, to handle the case $z=z'$ we apply a similar argument of analytical
continuation, using the fact that the expressions for the kernels are locally
holomorphic functions in $(z,z')$ \qed.
\enddemo

\head 3. The hypergeometric and gamma kernels  (second form)\endhead

Recall the definition of the {\it Frobenius coordinates\/} of a nonempty
diagram $\la\in\Y$: these are the integers $p_1>\dots> p_d\ge0$, $q_1>\dots>
q_d\ge0$, where $d$ is the number of boxes on the main diagonal of $\la$ and
$$
p_i=\la_i-i, \quad q_i=\la'_i-i, \qquad i=1,\dots,d.
$$
Any collection of integers $p_1>\dots> p_d\ge0$, $q_1>\dots>
q_d\ge0$ corresponds to a Young diagram. The transposition
$\la\mapsto\la'$ corresponds to interchanging $p_i\leftrightarrow
q_i$. In terms of Frobenius coordinates, the expression for the
z--measure, see Definition 1.4, can be rewritten as follows
$$
\gathered
M_{z,z',\xi}(\la)=(1-\xi)^{zz'}\,\xi^{|\la|}\,(zz')^d\\
\times \prod_{i=1}^d
(z+1)_{p_i}(z'+1)_{p_i}(-z+1)_{q_i}(-z'+1)_{q_i} \cdot
\left(\frac{\dim\la}{|\la|!}\right)^2,
\endgathered
$$
where
$$
\frac{\dim\la}{|\la|!}=\frac{\prod_{1\le i<j\le
d}(p_i-p_j)(q_i-q_j)}{\prod_{1\le i,j\le d}(p_i+q_j+1)\prod_{1\le i\le
d}p_i!q_i!}
$$

To any diagram $\la\in\Y$ we assign a finite subset $X(\la)\subset\Z'$:
$$
\gather X(\la)=X_+(\la)\sqcup X_-(\la),\\
X_+(\la)=\{\tp_1,\dots,\tp_d\}\subset\Z'_+, \qquad
X_-(\la)=\{-\tq_1,\dots,-\tq_d\}\subset\Z'_-,
\endgather
$$
where
$$
\tp_i=p_i+\tfrac12, \qquad \tq_i=q_i+\tfrac12, \qquad i=1,\dots,
d,
$$
are the {\it modified\/} Frobenius coordinates of $\la$ and
$$
\Z'_-=\{\dots,-\tfrac52,-\tfrac32,-\tfrac12\},  \qquad
\Z'_+=\{\tfrac12,\tfrac32,\tfrac52,\dots\}.
$$
By convention, $X(\varnothing)=\varnothing$. Note that $\la$ is uniquely
determined by $X(\la)$, so that the correspondence $\la\mapsto X(\la)$ is an
embedding of $\Y$ into the space $\{0,1\}^{\Z'}$.

\proclaim{Proposition 3.1} The correspondence $\la\mapsto X(\la)$,
defined above, and the correspondence $\la\mapsto \unX(\la)$,
which was defined at the beginning of \S2, are related to each
other as follows. For any $\la\in\Y$,
$$
X(\la)=\unX(\la)\;\triangle\;\Z'_-, \qquad \unX(\la)=X(\la)\;\triangle\;\Z'_-,
$$
where the symbol $\triangle$ denotes the symmetric difference of two sets.
\endproclaim

\demo{Proof} This can be proved using a simple geometric argument,
cf. Borodin--Olshanski \cite{BO4, \S4}. Notice that the claim is
equivalent to the classical Frobenius lemma, see Macdonald
\cite{Ma, Example I.1.15 (a)}. \footnote{This claim was exploited
in Borodin--Okounkov--Olshanski \cite{BOO, \tht{1.2}}. In that
paper, $\unX(\la)$ and $X(\la)$ were denoted as $\Cal D(\la)$ and
$\operatorname{Fr}(\la)$, respectively.}\qed
\enddemo

In terms of binary sequences, the claim of Proposition 3.1 can be
restated as follows. Let $a\mapsto a^\circ$ denote the involutive
homeomorphism of the space $\{0,1\}^{\Z'}$ which applies the
transposition $0\leftrightarrow1$ to all digits indexed by
negative semi--integers. Then we have $X(\la)=(\unX(\la))^\circ$,
$\unX(\la)=(X(\la))^\circ$.

Let $P_{z,z',\xi}$ be the push--forward of the measure
$M_{z,z',\xi}$ under the embedding
$\Y\hookrightarrow\{0,1\}^{\Z'}$ defined by the correspondence
$\la\mapsto X(\la)$. Then, by Proposition 3.1, $P_{z,z',\xi}$
coincides with image of the measure $\unP_{z,z',\xi}$ under the
involution $a\mapsto a^\circ$. We aim to write down  the
correlation functions of $P_{z,z',\xi}$.

\proclaim{Theorem 3.2} Let $(z,z')$ be as in Theorem 2.2. The
correlation functions of the measure $P_{z,z',\xi}$ have
determinantal form
$$
\rho_m(x_1,\dots,x_m\mid P_{z,z',\xi})=\det \Sb 1\le i,j\le m \endSb
[K(x_i,x_j)], \qquad m=1,2,\dots,
$$
where the kernel
$$
K(x,y)=K(x,y\mid z,z',\xi)
$$
on $\Z'\times \Z'$ is defined by the following formulas depending
on the sign of $x$ and $y$.

$\bullet$ For $x>0$, $y>0$\/{\rm:}
$$
\dfrac{P(x\mid z,z',\xi)\,Q(y\mid z,z',\xi) -Q(x\mid z,z',\xi)\,P(y\mid
z,z',\xi)}{x-y}\,.
$$

$\bullet$ For $x>0$, $y>0$\/{\rm:}
$$
\dfrac{P(x\mid z,z',\xi)\,P(-y\mid -z,-z',\xi) +Q(x\mid z,z',\xi)\,Q(-y\mid
-z,-z',\xi)}{x-y}\,.
$$

$\bullet$ For $x<0$, $y>0$\/{\rm:}
$$
\dfrac{P(-x\mid -z,-z',\xi)\,P(y\mid z,z',\xi) +Q(-x\mid -z,-z',\xi)\,Q(y\mid
z,z',\xi)}{x-y}\,.
$$

$\bullet$ For $x<0$, $y>0$\/{\rm:}
$$
\dfrac{P(-x\mid -z,-z',\xi)\,Q(-y\mid -z,-z',\xi) -Q(-x\mid
-z,-z',\xi)\,P(-y\mid -z,-z',\xi)}{-x+y}\,.
$$
Here $P$ and $Q$ are the functions introduced in Theorem 2.2.
\endproclaim

\demo{Comments} 1. Notice that $K(y,x)=\sgn(x)\sgn(y)K(x,y)$.

2. The indeterminacy $0/0$ on the diagonal $x=y$ is removed by making use of
the l'Hospital rule.
\enddemo

\demo{Proof} See Borodin--Olshanski \cite{BO2, Theorem 3.3}. \qed
\enddemo

Actually, Theorem 3.3 in \cite{BO2} contains a stronger claim (see
Theorem 3.4 below). In order to state it, we introduce a kernel
$A$ on $\Z'_+\times\Z'_-$ by
$$
\gather A(x,y\mid z,z',\xi) =\frac{\xi^{(x-y)/2}\sqrt{\sin(\pi
z)\sin(\pi z')}}{\pi} \\
\times \frac{\sqrt{\Ga(z+x+\tfrac12)\Ga(z'+x+\tfrac12)}}{\Ga(x+\tfrac12)} \cdot
\frac{\sqrt{\Ga(-z-y+\tfrac12)\Ga(-z'-y+\tfrac12)}}{\Ga(-y+\tfrac12)} \cdot
\frac1{x-y}\,,\\
x\in\Z'_+, \quad y\in\Z'_-\,.
\endgather
$$

\proclaim{Proposition 3.3} The kernel $A(x,y\mid z,z',\xi)$ is of trace class,
i.e., the corresponding operator $\ell^2(\Z'_-)\to\ell^2(\Z'_+)$ is of trace
class.
\endproclaim

\demo{Proof} First of all, note that the denominator $x-y$ does
not vanish, because $x-y\ge1$. Observe that
$$
\sum_{x\in\Z'_+,\, y\in\Z'_-}|A(x,y\mid z,z',\xi)|<\infty.
$$
Indeed, in the expression for the kernel, the ratios of gamma factors have at
most polynomial growth,
$$
\gathered
\frac{\sqrt{\Ga(z+x+\tfrac12)\Ga(z'+x+\tfrac12)}}{\Ga(x+\tfrac12)}\;\sim\;
x^{(z+z')/2}, \qquad x\to+\infty\\
\frac{\sqrt{\Ga(-z-y+\tfrac12)\Ga(-z'-y+\tfrac12)}}{\Ga(-y+\tfrac12)}\;\sim\;
|y|^{-(z+z')/2}, \qquad y\to-\infty
\endgathered \tag3.1
$$
while $\xi^{(x-y)/2}=\xi^{(x+|y|)/2}$ has an exponential decay.

Now the claim follows from a well--known sufficient condition: an
infinite matrix $A=[A_{ij}]$ is of trace class if the sum
$\sum|A_{ij}|$ is finite. Here is a simple argument that justifies
the sufficiency.

It is enough to show that $\Vert A \Vert_1\le \sum |A_{ij}|$,
where $\Vert A \Vert_1$ is the trace norm. We have
$$
\Vert A \Vert_1=\sup_B|\tr(AB)|,
$$
where $B$ ranges over the set of all (say, finite--dimensional)
matrices with $\Vert B\Vert\le1$, and $\Vert B\Vert$ is the
ordinary norm. But
$$
|\tr(AB)|=\big|\sum_{i,j}A_{ij} B_{ji}\big|\le \sum_{i,j}|A_{ij}|
\cdot| B_{ji}| \le \sum_{i,j}|A_{ij}|,
$$
because $\Vert B\Vert\le1$ implies $|B_{ji}|\le1$.
 \qed
\enddemo

Next, introduce a kernel $L$ on $\Z'\times\Z'$ by
$$
L(x,y\mid z,z',\xi)=\cases 0, & x>0,\;y>0 \\ A(x,y\mid z,z',\xi), & x>0, \;
y<0\\
-A(y,x\mid z,z',\xi), & x<0, \; y>0\\ 0, & x<0, \; y<0. \endcases
\tag3.2
$$
Let $L$ be the operator in the Hilbert space $\ell^2(\Z')$ defined by this
kernel. By Proposition 3.3, $L$ is of trace class, so that $\det(1+L)$ makes
sense.

It is readily checked (\cite{BO2, Proposition 3.1}) that
$$
M_{z,z',\xi}(\la)=\dfrac{\det\limits\Sb x,y\in X(\la)\endSb [L(x,y\mid
z,z',\xi)]}{\det(1+L)}\,, \qquad \la\in\Y.
$$
By a general claim (see \cite{BO2, \S2}), this implies that
$$
\rho_m(x_1,\dots,x_m\mid P_{z,z',\xi})=\det \Sb 1\le i,j\le m \endSb
\left[\frac{L}{1+L}(x_i,x_j)\right].
$$

\proclaim{Theorem 3.4} Let $L$ be the operator in $\ell^2(\Z')$ with kernel
$L(x,y\mid z,z',\xi)$. The kernel $K(x,y\mid z,z',\xi)$ is precisely the matrix
of the operator $\dfrac{L}{1+L}$.
\endproclaim

\demo{Proof} See Borodin--Olshanski \cite{BO2, Theorem 3.3}. \qed
\enddemo

The next claim is a counterpart of Theorem 2.3.

\proclaim{Theorem 3.5}  As $\xi\nearrow1$, the measures $P_{z,z',\xi}$ weakly
converge to a probability measure $P^\gam_{z,z'}$ on $\{0,1\}^{\Z'}$. The
correlation functions of the limit measure have determinantal form,
$$
\gather \rho_m(x_1,\dots,x_m\mid P^\gam_{z,z'})=\det_{1\le i,j\le
m}[K^\gam(x_i,x_j\mid z,z')], \\
m=1,2,\dots, \quad x_1,\dots,x_m\in\Z',
\endgather
$$
where the kernel $K^\gam(x,y\mid z,z')$ on $\Z'\times\Z'$, which
is equal to the pointwise limit of the kernel $K(x,y\mid
z,z',\xi)$ as $\xi\nearrow 1$, is given by the following formulas
depending on the signs of the arguments $x,y$.

$\bullet$ For $x>0$, $y>0$, the kernel is given by same expression as in
Theorem 2.3\/{\rm:}
$$
\gather \frac{\sin(\pi z)\sin(\pi z')}{\pi\sin(\pi(z-z'))}
\,\left\{\Ga(z+x+\tfrac12)\Ga(z'+x+\tfrac12)
\Ga(z+y+\tfrac12)\Ga(z'+y+\tfrac12)\right\}^{-1/2}\\ \times
\frac{\Ga(z+x+\tfrac12)\Ga(z'+y+\tfrac12)
-\Ga(z'+x+\tfrac12)\Ga(z+y+\tfrac12)}{x-y}
\endgather
$$

$\bullet$ For $x>0$, $y<0$\/{\rm:}
$$
\gather \frac{\sqrt{\sin(\pi z)\sin(\pi z')}}{\pi\sin(\pi(z-z'))}
\,\left\{\Ga(z+x+\tfrac12)\Ga(z'+x+\tfrac12)
\Ga(-z-y+\tfrac12)\Ga(-z'-y+\tfrac12)\right\}^{-1/2}\\
\times \frac{\sin(\pi z)\Ga(z+x+\tfrac12)\Ga(-z-y+\tfrac12)-\sin(\pi
z')\Ga(z'+x+\tfrac12)\Ga(-z'-y+\tfrac12)}{x-y}
\endgather
$$

$\bullet$ For $x<0$, $y>0$\/{\rm:}
$$
\gather \frac{\sqrt{\sin(\pi z)\sin(\pi z')}}{\pi\sin(\pi(z-z'))}
\,\left\{\Ga(-z-x+\tfrac12)\Ga(-z'-x+\tfrac12)
\Ga(z+y+\tfrac12)\Ga(z'+y+\tfrac12)\right\}^{-1/2}\\
\times \frac{\sin(\pi z)\Ga(-z-x+\tfrac12) \Ga(z+y+\tfrac12)-\sin(\pi z')
\Ga(-z'-x+\tfrac12)\Ga(z'+y+\tfrac12)}{x-y}
\endgather
$$

$\bullet$ For $x<0$, $y<0$\/{\rm:}
$$
\gather \frac{\sin(\pi z)\sin(\pi z')}{\pi\sin(\pi(z-z'))}
\,\left\{\Ga(-z-x+\tfrac12)\Ga(-z'-x+\tfrac12)
\Ga(-z-y+\tfrac12)\Ga(-z'-y+\tfrac12)\right\}^{-1/2}\\
\times \frac{\Ga(-z-x+\tfrac12)\Ga(-z'-y+\tfrac12)
-\Ga(-z'-x+\tfrac12)\Ga(-z-y+\tfrac12)}{x-y}
\endgather
$$
\endproclaim

\demo{Proof} The case $x,y>0$  was proved in Theorem 2.3. This immediately
implies the case $x,y<0$, because of an obvious symmetry of the formulas of
Theorem 3.2 (changing the signs of $x,y$ is equivalent to changing the signs of
$z,z'$). In the remaining two cases we argue just as in the proof of Theorem
2.3. \qed
\enddemo

\head 4. The relation between two forms of kernels \endhead

Our next goal is to describe a relation between the two types of
the discrete hypergeometric kernel $\unK(x,y\mid z,z',\xi)$ and
$K(x,y\mid z,z',\xi)$ and, similarly, between the two types of the
gamma kernel $\unK^\gam(x,y\mid z,z')$ and $K^\gam(x,y\mid z,z')$.

Given an arbitrary kernel $K(x,y)$ on $\Z'\times\Z'$, we assign to
it another kernel,
$$
K^\circ(x,y)=\cases K(x,y), & x>0,\\
\de_{xy}- K(x,y), & x<0,\endcases
$$
where $\de_{xy}$ is the Kronecker symbol. Slightly more generally, given an
arbitrary map $\epsi: \Z'\to\R^*$, we set
$$
K^{\circ,\epsi}(x,y)=\epsi(x) K^\circ(x,y)\epsi(y)^{-1}.
$$

\proclaim{Proposition 4.1} Let $P$ be a probability measure on
$\{0,1\}^{\Z'}$ and $P^\circ$ be its image under the involutive
homeomorphism $a\mapsto a^\circ$ of the space $\{0,1\}^{\Z'}$,
introduced after Proposition 3.1. Assume that the correlation
functions of $P$ have determinantal form with a certain kernel
$K(x,y)$,
$$
\rho_m(x_1,\dots,x_m\mid P)=\det \Sb 1\le i,j\le m\endSb
[K(x_i,x_j)], \qquad m=1,2,\dots\,.
$$
Then the correlation functions of the measure $P^\circ$ also have a similar
determinantal form, with the kernel $K^\circ(x,y)$ as defined above or, equally
well, with the kernel $K^{\circ,\epsi}(x,y)$, where the map $\epsi:\Z'\to\R^*$
may be chosen arbitrarily,
$$
\gathered \rho_m(x_1,\dots,x_m\mid P^\circ)=\det \Sb 1\le i,j\le
m\endSb [K^\circ(x_i,x_j)]=\det \Sb 1\le i,j\le m\endSb
[K^{\circ,\epsi}(x_i,x_j)],\\
 m=1,2,\dots\,.
\endgathered
$$
\endproclaim
\demo{Proof} The factor $\epsi(\,\cdot\,)$ does not affect the values of
determinants in right--hand side of the above formula, so that we may take
$\epsi(\,\cdot\,)\equiv1$. Then the result is obtained by applying the
inclusion/exclusion principle, see Proposition A.8 in
Borodin--Okounkov--Olshanski \cite{BOO}. \qed
\enddemo

\proclaim{Theorem 4.2} The kernels $\unK(x,y\mid z,z',\xi)$ and $K(x,y\mid
z,z',\xi)$, introduced in \S2 and \S3, respectively, are related to each other
by the transformation $K\mapsto K^{\circ,\epsi}$, where
$$
\epsi(x)=\cases 1, & x\in\Z'_+, \\ (-1)^k, &
x=-(k+\tfrac12)\in\Z'_-, \quad k=0,1,2,\dots\,.\endcases
$$
\endproclaim

\demo{Comments} 1. Since the kernels in question are associated with the
measures $\unP_{z,z',\xi}$ and $P_{z,z',\xi}$, which are related to each other
by the involution, the claim of the proposition is not surprising, in view of
Proposition 4.1. The point is the explicit form of the factor
$\epsi(\,\cdot\,)$.

2. The claim of the theorem generalizes Lemma 2.5 in
Borodin--Okounkov--Olshanski \cite{BOO}.
\enddemo

Before giving a proof let us state a corollary.

\proclaim{Corollary 4.3}  Theorem 2.3 and Theorem 3.2 are equivalent.
\endproclaim

\demo{Proof} Indeed, this follows from Proposition 4.1 and Theorem
4.2. \qed
\enddemo

\demo{Proof of Theorem 4.2} (a) Let us check the desired relation
between $\unK(x,y\mid z,z',\xi)$ and $K(x,y\mid z,z',\xi)$ for an
arbitrary couple $x,y$ outside the diagonal $x=y\in\Z'$. The
classical transformation formula \cite{Er1, 2.8(19)} implies
$$
\gather \frac1{\Ga(c)}\,F(a,b;c;\tfrac\xi{\xi-1})\bigg|_{c=-k}
=(-1)^{k+1}\xi^{k+1}(1-\xi)^{a+b-1}(a)_{k+1}(b)_{k+1}\\
\times \frac1{\Ga(k+2)}\, F(1-a,1-b;k+2;\tfrac\xi{\xi-1})\\
\text{ for any $a,b\in\C$ and any $k=-1,0,1,2,\dots$.}
\endgather
$$
{}From this we derive
$$
\gathered
P(x\mid z,z',\xi)=(-1)^{x-1/2}Q(-x\mid -z,-z',\xi), \qquad x\in\Z'_-\\
Q(x\mid z,z',\xi)=(-1)^{x+1/2}P(-x\mid -z,-z',\xi), \qquad x\in\Z'_-.
\endgathered  \tag4.1
$$
This readily implies the relation in question.

(b) Consider now the case $x=y\in\Z'$. We have to prove that
$$
K(x,x\mid z,z',\xi)=1-\unK(x,x\mid z,z',\xi), \qquad x\in\Z'_-.
$$
First, we will prove that
$$
\frac{d}{d\xi}\,K(x,x\mid z,z',\xi) =-\frac{d}{d\xi}\,\unK(x,x\mid z,z',\xi),
\qquad x\in\Z'_-.
$$
By virtue of Proposition 4.5 below, this is equivalent to
$$
P(x\mid z,z',\xi)Q(x\mid z,z',\xi) =-P(-x\mid -z,-z',\xi)Q(-x\mid -z,-z',\xi),
\qquad x\in\Z'_-,
$$
which in turn follows from formulas \tht{4.1} above.

(c) To conclude the proof it suffices to prove that
$$
\lim_{\xi\nearrow1}K(x,x\mid z,z',\xi) =\lim_{\xi\nearrow1}(1-\unK(x,x\mid
z,z',\xi)), \qquad x\in\Z'_-.
$$
By virtue of Theorem 2.3, this means
$$
\unK^\gam(x,x\mid z,z')+\unK^\gam(-x,-x\mid -z,-z')=1, \qquad x\in\Z'_-\,.
$$
Using Comment 3 to Theorem 2.3 we reduce this to
$$
\multline \frac{\sin(\pi z)\sin(\pi z')}{\pi\sin(\pi(z-z'))}\,
\big\{\psi(z+x+\tfrac12)-\psi(z'+x+\tfrac12)\\
-\psi(-z-x+\tfrac12)+\psi(-z'-x+\tfrac12)\big\}=1, \endmultline
$$
which is verified using a well--known relation for the
$\psi$--function \cite{Er1, 1.7.1(8)}:
$$
\psi(a)-\psi(1-a)=-\pi \operatorname{ctg}(\pi a).\qed
$$
\enddemo

The counterpart of Theorem 4.2 for the gamma kernels is

\proclaim{Theorem 4.4} The kernels $\unK^\gam(x,y\mid z,z')$ and
$K^\gam(x,y\mid z,z')$, introduced in \S2 and \S3, respectively, are related to
each other by the transformation $K\mapsto K^{\circ,\epsi}$, where
$$
\epsi(x)=\cases 1, & x\in\Z'_+ \\ (-1)^k, & x=-(k+\tfrac12)\in\Z'_-, \quad
k=0,1,2,\dots\endcases
$$
\endproclaim

\demo{Proof} This follows from Theorem 4.2 if we pass to the limit
as $\xi\nearrow 1$. On the other hand, this can be readily checked
directly, because the crucial step, the coincidence of both
kernels for $x=y\in\Z'_-$, was already verified in the proof of
Theorem 4.2. \qed
\enddemo

The next result, which we have just used in the proof of Theorem 4.2, is also
of independent interest. It is a generalization of the differentiation formula
for the discrete Bessel kernel, see Borodin--Okounkov--Olshanski \cite{BOO,
\tht{2.11} and below}.

\proclaim{Proposition 4.5} We have
$$
\frac{d}{d\xi}\,\unK(x,y\mid z,z',\xi) =\frac1{2\xi}\,\left(P(x\mid
z,z',\xi)Q(y\mid z,z',\xi)+Q(x\mid z,z',\xi)P(y\mid z,z',\xi)\right).
$$
\endproclaim

\demo{Proof} This can be directly verified by making use of the
differentiation formulas
$$
\gather \frac{d}{d\xi}\, P(x\mid
z,z',\xi)=\left(\frac{x}{2\xi}-\frac{z+z'}{2(1-\xi)}\right)P(x\mid
z,z',\xi)-\frac{(zz')^{1/2}}{\xi^{1/2}(1-\xi)}Q(x\mid z,z',\xi), \\
\frac{d}{d\xi}\, Q(x\mid
z,z',\xi)=\left(-\,\frac{x}{2\xi}+\frac{z+z'}{2(1-\xi)}\right)Q(x\mid
z,z',\xi)+\frac{(zz')^{1/2}}{\xi^{1/2}(1-\xi)}P(x\mid z,z',\xi).
\endgather
$$
To check these formulas we use the following differentiation
formulas for the Gauss hypergeometric function, which can be
derived from \cite{Er1, 2.8 (20), (27)}:
$$
\gather \frac{d}{d\xi}\left(\frac{F(a,b;c;\frac\xi{\xi-1})}{\Ga(c)}\right) =
-\,\frac{ab}{(1-\xi)^2}\, \frac{F(a+1,b+1;c+1;\frac\xi{\xi-1})}{\Ga(c+1)}\\
=\frac1\xi\,\frac{F(a-1,b-1;c-1;\frac\xi{\xi-1})}{\Ga(c-1)} -
\left(\frac{a+b-1}{1-\xi}
+\frac{c-1}{\xi}\right)\,\frac{F(a,b;c;\frac\xi{\xi-1})}{\Ga(c)}\,.
\qed
\endgather
$$
\enddemo

\head 5. The projection property \endhead

Let $H=H_+\oplus H_-$ be a Hilbert space decomposed into a direct sum of two
subspaces. According to this decomposition we will write operators in $H$ in
$2\times2$ block form. Let $A: H_-\to H_+$ be a bounded operator and let
$$
L=\bmatrix 0 & A\\ -A^* & 0 \endbmatrix.
$$
This is a bounded operator in $H$. Notice that $1+L$ is invertible. Indeed this
follows from the fact that
$$
(1+L)^*(1+L)=\bmatrix 1+AA^* & 0\\ 0 & 1+A^*A \endbmatrix\;\ge\; \bmatrix 1 &
0\\ 0 & 1 \endbmatrix.
$$
Set $K=L(1+L)^{-1}$ and write $K$ in the block form,
$$
K=\bmatrix a & b \\ c & d \endbmatrix.
$$
Next, set
$$
K^\circ=\bmatrix a & b \\ -c & 1-d \endbmatrix, \qquad {}^\circ K=\bmatrix 1-a &
-b \\
c & d
\endbmatrix.
$$

\proclaim{Proposition 5.1} The operators $K^\circ$ and ${}^\circ K$ as defined
above are orthogonal projections onto the subspaces
$$
H^\circ=\{Ah_-\oplus h_-\mid h_-\in H_-\}, \qquad  {}^\circ
H=\{h_+\oplus(-A^*)h_+\mid h_+\in H_+\},
$$
which are essentially the graphs of the operators $A$ and $-A^*$, respectively.
We have $K^\circ\cdot{}^\circ K={}^\circ K\cdot K^\circ=0$ and
$K^\circ+{}^\circ K =1$.
\endproclaim

\demo{Proof} The latter equality is immediate from the definition of $K^\circ$
and ${}^\circ K$.

Obviously, $H^\circ$ and ${}^\circ H$ are closed subspaces, orthogonal to each
other. Moreover, as is well known, their sum is the whole $H$. (Indeed, it
suffices to check that any $f\in H_+$ can be written as a sum of vectors from
$H^\circ$ and ${}^\circ H$. This means that
$$
h_--A^*h_+=0, \qquad Ah_-+h_+=f,
$$
which is reduced to $(1+AA^*)h_+=f$. But the latter equation is solvable,
because $1+AA^*$ is invertible.)

Next, one can directly verify that
$$
\gather
a=(1+AA^*)^{-1}AA^*=AA^*(1+AA^*)^{-1}\\
b=(1+AA^*)^{-1}A=A(1+A^*A)^{-1}\\
c=-(1+A^*A)^{-1}A^*=-A^*(1+AA^*)^{-1}\\
d=(1+A^*A)^{-1}A^*A=A^*A(1+A^*A)^{-1}\,.
\endgather
$$
Using these explicit expressions for the blocks $a,b,c,d$ one can readily check
that the operator $K^\circ$ is the identity on $H^\circ$ and zero on ${}^\circ
H$. Similarly, the operator ${}^\circ K$ is the identity on ${}^\circ H$ and
zero on $H^\circ$. This concludes the proof. \qed
\enddemo

\example{Remark 5.2} The claim of Proposition 5.1 remains true under weaker
assumptions. Namely, $A$ may be an unbounded, closed operator with dense
domain. \qed
\endexample

\proclaim{Theorem 5.3} The discrete hypergeometric kernel $\unK(x,y\mid
z,z',\xi)$ on $\Z'\times\Z'$, as defined in \S2, is a projection kernel. That
is, it corresponds to an orthogonal projection operator in the Hilbert space
$\ell^2(\Z')$.
\endproclaim

\demo{Proof} Take $H=\ell^2(\Z')$, $H_+=\ell^2(\Z'_+)$,
$H_-=\ell^2(\Z'_-)$, and let $K$ be the operator in $H$ defined by
the kernel $K(x,y\mid z,z',\xi)$. By Theorem 3.4, $K=L(1+L)^{-1}$,
where $L$ has the form $\bmatrix 0 & A\\ -A^* & 0\endbmatrix$ with
a certain bounded operator $A$ (recall that the kernel $A(x,y)$ is
real, so that the adjoint operator $A^*$ is given by the
transposed kernel). Let $\unK$ be the operator given by the kernel
$\unK(x,y\mid z,z',\xi)$. By Theorem 4.2, $\unK=\epsi K^\circ
\epsi^{-1}$, where $\epsi$ is a diagonal matrix with $\pm1$'s on
the diagonal. By Proposition 5.1, the operator $K^\circ$ is an
orthoprojection. Therefore, $\unK$ is an orthoprojection, too.
\qed
\enddemo

We would like to prove a similar claim for the gamma kernel
$\unK^\gam(x,y\mid z,z')$. By Theorem 2.3, it is the pointwise
limit (as $\xi\nearrow1$) of the hypergeometric kernels
$\unK(x,y\mid z,z',\xi)$, which are projection kernels by virtue
of Theorem 5.3. That is, the operator defined by the gamma
kernel is a weak limit of orthoprojections.  However, the
projection property is not stable under limit transitions in the
weak operator topology. Indeed, one can obtain any selfadjoint
operator with norm $\le1$
 as a weak limit of orthoprojections in an
infinite--dimensional Hilbert space. It would be nice to
strengthen Theorem 2.3 by proving that the kernels (or rather the
corresponding operators) actually converge in the {\it strong\/}
operator topology: this would suffice to conclude that the limit
kernel inherits the projection property. However, to derive the
strong convergence directly from the formulas, as we have done for
the weak convergence, does not seem to be easy.

Below we present a simple argument, which proves the strong convergence in a
roundabout way, under an additional restriction on the parameters $z,z'$. The
idea is to prove an analog of Theorem 3.4. To do this we verify the strong
convergence of the ``$L$--operators'', whose kernels are much simpler than
those of the ``$K$--operators''. We will impose a restriction on the parameters
$z,z'$ to ensure the boundedness of the limit ``$L$--operator''.

Consider the kernel $A(x,y\mid z,z',\xi)$ on $\Z'_+\times\Z'_-$ introduced in
\S3. As $\xi\nearrow1$, the factor $\xi^{(x-y)/2}$ tends to 1, so that the
kernel pointwise converges to the kernel
$$
\gather A(x,y\mid z,z'): =\frac{\sqrt{\sin(\pi z)\sin(\pi z')}}{\pi} \\
\times \frac{\sqrt{\Ga(z+x+\tfrac12)\Ga(z'+x+\tfrac12)}}{\Ga(x+\tfrac12)} \cdot
\frac{\sqrt{\Ga(-z-y+\tfrac12)\Ga(-z'-y+\tfrac12)}}{\Ga(-y+\tfrac12)} \cdot
\frac1{x-y}\,,\\
x\in\Z'_+, \quad y\in\Z'_-\,.
\endgather
$$

\proclaim{Proposition 5.4} Assume $|z+z'|<1$. Then the operator
$A:\ell^2(\Z'_-)\to\ell^2(\Z'_+)$ with the kernel $A(x,y\mid
z,z')$ is bounded.

Furthermore, let $A_\xi$ stand for the operator with the kernel $A(x,y\mid
z,z',\xi)$. As $\xi\nearrow1$, we have $A_\xi\to A$ and $A^*_\xi\to A^*$ in the
strong operator topology.
\endproclaim

Recall that $(z,z')$ is subject to one of the conditions (i),
(ii) of Proposition 1.8. This implies, in particular, that
$z+z'$ is real. If $(z,z')$ satisfies condition (i) then the
additional restriction $|z+z'|<1$  means that $z=\overline{z'}$
lies in the strip $\Re(\,\cdot\,)<1/2$.

\demo{Proof} The second claim easily follows from the first one.
Indeed, we have $A_\xi=\Xi\, A\, \Xi$, where
$\Xi=\operatorname{diag(1,\xi,\xi^2,\dots)}$. Observe that
$\Vert \Xi\Vert =1$, $\Xi$ converges to the identity operator in
the strong operator topology. Next, we have $\Vert \Xi\Vert=1$
for any $\xi$. Since operator multiplication is a jointly
strongly continuous operation on bounded sets, the strong
convergence $A_\xi\to A$ and $A_\xi^*\to A^*$ follows.

Let us prove the first claim. Let $f$ and $g$ range over the unit
balls of the Hilbert spaces $\ell^2(\Z'_+)$ and $\ell^2(\Z'_-)$,
respectively. We have
$$
\Vert A\Vert=\sup_{f,g}|(Af,g)|
\le\sup_{f,g}\left(\sum_{x\in\Z'_+}\sum_{y\in\Z'_-} |A(x,y\mid
z,z')|\cdot|f(x)|\cdot|g(y)|\right).
$$
By virtue of \tht{3.1}, in order to prove that this quantity is
finite, it suffices to prove that
$$
\sup_{f,g}\left(\sum_{x\in\Z'_+}\sum_{y\in\Z'_-}
\left(\frac{x}{|y|}\right)^{\frac{z+z'}2}\cdot\frac1{x+|y|}\cdot
|f(x)|\cdot|g(y)|\right)\;<\;+\infty.
$$
It is convenient to rewrite this as
$$
\sup_{f,g}\left(\sum_{x\in\Z'_+}\sum_{y\in\Z'_+}
\left(\frac{x}{y}\right)^{\frac{z+z'}2}
\cdot\frac1{x+y}\cdot|f(x)|\cdot|g(y)|\right)\;<\;+\infty.
$$
Here we assume that both $f$ and $g$ range over the unit ball of
$\ell^2(\Z'_+)$.

Next, we may replace the sums over $\Z'_+$ by the integrals over $\R_+$ with
respect to Lebesgue measure (assuming that $f$ and $g$ range over the unit ball
of $L^2(\R_+, dx)$). Indeed, this will only strengthen the claim. The resulting
claim is equivalent to the boundedness of the operator in $L^2(\R_+, dx)$ with
the kernel
$$
\left(\frac{x}{y}\right)^{\frac{z+z'}2}\cdot\frac1{x+y}\,.
$$
It is not hard to show that this integral operator is bounded if
and only if $|z+z'|<1$, see Olshanski \cite{Ol1}. \qed
\enddemo

Similarly to \tht{3.2}, using the kernel $A(x,y\mid z,z')$ we
construct another kernel on $\Z'\times\Z'$ by
$$
L(x,y\mid z,z')=\cases 0, & x>0,\;y>0 \\ A(x,y\mid z,z'), & x>0, \; y<0\\
-A(y,x\mid z,z'), & x<0, \; y>0\\ 0, & x<0, \; y<0. \endcases
$$

The next result is the counterpart of Theorem 3.4.

\proclaim{Theorem 5.5} Let $L: \ell^2(\Z')\to\ell^2(\Z')$ be the
operator with kernel $L(x,y\mid z,z')$. Assume $|z+z'|<1$. Then
the kernel $K^\gam(x,y\mid z,z')$ is precisely the matrix of the
operator $\dfrac{L}{1+L}$.
\endproclaim

\demo{Proof} Let $L_\xi$ denote the operator with kernel
$L(x,y\mid z,z',\xi)$. We claim that $L_\xi(1+L_\xi)^{-1}$
strongly converges to $L(1+L)^{-1}$. To check this we use a
standard argument. Write the formal identity
$$
(1+L_\xi)^{-1}-(1+L)^{-1}=(1+L_\xi)^{-1}(L-L_\xi)(1+L)^{-1}.
$$
Since $(1+L^*_\xi)(1+L_\xi)\ge1$ (see the beginning of the
section), we have $\Vert (1+L_\xi)^{-1}\Vert\le 1$. Next, the
operators $L_\xi$ are uniformly bounded and $L_\xi\to L$
strongly: this follows from Proposition 5.4 (here we use the
assumption $|z+z'|<1$). Therefore, the product in the
right--hand side strongly converges to 0.

Since the kernel of $L_\xi(1+L_\xi)^{-1}$ is $K(x,y\mid
z,z',\xi)$ (Theorem 3.4), the latter kernel strongly converges
to the kernel of $L(1+L)^{-1}$. On the other hand, we already
know (Theorem 3.5) that $K(x,y\mid z,z',\xi)$ pointwise
converges to $K^\gam(x,y\mid z,z')$. We conclude that
$K^\gam(x,y\mid z,z')$ is the kernel of $L(1+L)^{-1}$. \qed
\enddemo

\proclaim{Theorem 5.6} Assume $|z+z'|<1$. The gamma kernel
$\unK^\gam(x,y\mid z,z')$ on $\Z'\times\Z'$, as defined in \S2, is
a projection kernel.
\endproclaim

\demo{Proof} We argue precisely as in the proof of Theorem 5.3, with reference
to Theorem 4.4 instead of Theorem 4.2. \qed
\enddemo

We conjecture that the claim of Theorem 5.6 holds without the
restriction $|z+z'|<1$.

\head 6. The tail kernel \endhead

Let us study the asymptotics of the process $\unP_{z,z'}^\gam$
(equivalently, of $P^\gam_{z,z'}$) near $+\infty$. In order to
find a suitable scaling, let us look at the first correlation
function (also called the {\it density function}). It is given by
the value of the correlation kernel on the diagonal, which was
written down in Comment 3 to Theorem 2.3 in terms of the psi
function $\psi(x)$. Near $+\infty$, the psi function behaves as
follows (\cite{Er1, 1.19(7)})
$$
\psi(x)=\log x-\frac1{2x}+O(x^{-2}).
$$
Substituting this into the expression for $\unK^\gam(x,x\mid z,z')$ we see that
the density function of the process $\unP_{z,z'}^\gam$ behaves as
$$
\frac{(z-z')\sin(\pi z)\sin(\pi z')}{\pi\sin(\pi(z-z'))}\cdot
x^{-1}, \qquad x\to+\infty. \tag 6.1
$$
This suggests that the scaling should have the form $x=e^{s_0+s}$, where
$s_0\to+\infty$, because then in the coordinate $s$, the density function will
be asymptotically constant. Notice that in the limit transition, the lattice
turns into the real line.

All statements of this section are made under the assumption that
$(z,z')$ satisfy one of the conditions (i), (ii) of Proposition
1.8.

\proclaim{Proposition 6.1} In the scaling limit $x=e^{s_0+s}$, where
$s_0\to+\infty$, the correlation functions of $\unP_{z,z'}^\gam$ converge, and
the limit functions have determinantal form with the kernel
$$
\unK^{\tail}(s,t\mid z,z') =\frac{\sin(\pi z)\sin(\pi
z')}{\pi\sin(\pi(z-z'))}\cdot \frac{\sinh
(\tfrac12(z-z')(s-t))}{\sinh(\tfrac12(s-t))}, \qquad s,t\in\R.
$$
\endproclaim

\demo{Proof} It suffices to examine the limit behavior of the
correlation kernel $\unK^\gam(x,y\mid z,z')$. Recall that if we
are given a correlation kernel $K(x,y)$ on a state space with
reference measure $dx$ then, under a transformation of the state
space, we have to look at the transformation of the expression
$\sqrt{dxdy}\,K(x,y)$, rather than of $K(x,y)$ itself. In our
situation, $x=\exp(s_0+s)$, $y=\exp(s_0+t)$, so that
$\sqrt{dxdy}=\sqrt{xy}\sqrt{dsdt}$. Using the well--known
asymptotics of the ratio of gamma functions (\cite{Er1, 1.18 (4)})
and the explicit expression of the kernel in question we find that
the limit
$$
\lim_{s_0\to+\infty}\left\{\sqrt{xy}\;\unK^\gam(x,y\mid
z,z')\bigm|_{x=\exp(s_0+s), \,y=\exp(s_0+t)}\right\}
$$
exists and equals $\unK^{\tail}(s,t\mid z,z')$. \qed
\enddemo

We call $\unK^{\tail}(s,t\mid z,z')$ the {\it tail kernel\/} with
parameters $z,z'$. It determines a translationally invariant point
process on $\R$. The tail kernel was obtained via a double limit
transition: first, from the discrete hypergeometric kernel to the
gamma kernel, and next, from the gamma kernel to the tail kernel.
The same result can be obtained in one step, as the following
proposition shows.

\proclaim{Proposition 6.2} Consider the discrete hypergeometric
kernel $\unK(x,y\mid z,z',\xi)$ on the lattice $\Z'$, and make the
change of variables $x=e^{s_0+s}$, $y=e^{s_0+t}$. Let
$s_0\to+\infty$ and $\xi\nearrow1$. Moreover, assume that
$(1-\xi)^{-1}$ grows faster than $e^{s_0}$; namely,
$$
e^{s_0}=O\left((1-\xi)^{-\epsi}\right)
$$
where $\epsi>0$ is small enough {\rm(}it suffices to assume that
$\epsi$ is smaller than $1-|\Re(z-z')|$; we recall that
$|\Re(z-z')|<1$, see Proposition 1.8{\rm)}. Then the scaling limit
of the kernel $\unK(x,y\mid z,z',\xi)$ is the tail kernel.
\endproclaim

\demo{Proof} First, we slightly revise the proof of Theorem 2.3.
Specifically, we cannot apply the trivial estimate \tht{2.1},
because the parameter $c=x+\frac12$ is no longer constant. Instead
of this we use the Euler integral representation of the Gauss
hypergeometric function in the form
$$
F(a,1-c+a;1-b+a;w^{-1})=\Ga(1-b+a)\, \int_0^1
\frac{u^{a-1}}{\Ga(a)}\,\frac{(1-u)^{-b}}{\Ga(1-b)}\,(1-uw^{-1})^{c+a-1}du
$$
(see \cite{Er1, 2.1.3 (10)}), where, as in Theorem 2.3,
$(a,b,c;w)$ is either
$$
\left(-z,\,-z',\,x+\tfrac 12;\,\tfrac\xi{\xi-1}\right)\quad\text{
 or  }\quad\left(-z+1,\,-z'+1,\,x+\tfrac32;\,\tfrac\xi{\xi-1}\right).
$$
(Notice that in \cite{Er1} the Euler integral representation is
given under restrictions on the parameters. However, these
restrictions are inessential, because, in our notation, the
expression $\frac{u^{a-1}}{\Ga(a)}\,\frac{(1-u)^{-b}}{\Ga(1-b)}$
makes sense as a distribution supported by $[0,1]$, for {\it
any\/} complex $a,b$.)

By our hypothesis, $(c+a-1)w^{-1}=O\left((1-\xi)^{1-\epsi}\right)$, whence
$$
(1-uw^{-1})^{c+a-1}=1+O\left((1-\xi)^{1-\epsi}\right)
$$
uniformly in $u\in[0,1]$. This gives
$$
F(a,1-c+a;1-b+a;w^{-1})=1+O\left((1-\xi)^{1-\epsi}\right)
$$
and likewise
$$
F(b,1-c+b;1-a+b;w^{-1})=1+O\left((1-\xi)^{1-\epsi}\right).
$$

Then we may continue the argument as in the proof of Theorem 2.3 and use the
same simple estimate for ratios of gamma functions as in the proof of
Proposition 6.1. \qed
\enddemo

The gamma kernel and the discrete hypergeometric kernel in the
second form (which corresponds to looking at the Frobenius
coordinates of Young diagrams) also have tail limits.

The density function of $P^\gam_{z,z'}$ has the asymptotics
$$
\frac{(z-z')\sin(\pi z)\sin(\pi z')}{\pi\sin(\pi(z-z'))}\cdot
|x|^{-1}, \qquad x\to \pm\infty,
$$
which immediately follows from the asymptotics \tht{6.1} of the
density function of $\unP^\gam_{z,z'}$. Indeed, $P^\gam_{z,z'}$
and $\unP^\gam_{z,z'}$ coincide on $\Z'_+$, and the change of sign
transformation of $P^\gam_{z,z'}$ is equivalent to changing the
signs of the parameters $z,z'$. Thus, it makes sense to consider
the scaling limit of $P^\gam_{z,z'}$ at both plus and minus
infinity.

\proclaim{Proposition 6.3} In the scaling limit $x=\pm e^{s_0+s}$,
where $s_0\to+\infty$, the correlation functions of
$P^\gam_{z,z'}$ converge, and the limit functions have
determinantal form with the kernel given by
$$
\align \bullet& \text{    For $x=e^{s_0+s}$, $y=e^{s_0+t}$, the
limit is the same as in Proposition 6.1:}\\
& \frac{\sin(\pi z)\sin(\pi z')}{\pi\sin(\pi(z-z'))}\cdot
\frac{\sinh (\tfrac12(z-z')(s-t))}{\sinh(\tfrac12(s-t))}\\
\bullet&\text{ For $x=e^{s_0+s}$, $y=-e^{s_0+t}$:}\\
&\frac{\sqrt{\sin(\pi z)\sin(\pi z')}}{\pi\sin(\pi(z-z'))}\cdot
\frac{\sin(\pi z)e^{\frac 12(z-z')(s-t)}-\sin(\pi z')e^{-\frac
12(z-z')(s-t)}}{2\cosh(\frac 12(s-t))}\\
\bullet&\text{ For $x=-e^{s_0+s}$, $y=e^{s_0+t}$:}\\
&\frac{\sqrt{\sin(\pi z)\sin(\pi z')}}{\pi\sin(\pi(z-z'))}\cdot
\frac{\sin(\pi z')e^{\frac 12(z-z')(s-t)}-\sin(\pi z)e^{-\frac
12(z-z')(s-t)}}{2\cosh(\frac 12(s-t))}\\
\bullet&\text{ For $x=-e^{s_0+s}$, $y=-e^{s_0+t}$ the kernel is
the same as for }\\ &\text{ $x=e^{s_0+s}$, $y=e^{s_0+t}$ (the
first case above).}
\endalign
$$
\endproclaim
We denote the resulting tail kernel in the second form by
$K^\tail(s,t\mid z,z')$. It defines a determinantal point process
on $\R\times \R$ which is invariant under simultaneous
translations $(s,t)\mapsto (s+\Delta, t+\Delta)$, $\Delta\in\R$.
This kernel appeared for the first time in \cite{Ol1, Proposition
4.1}, see also \cite{BO1, Theorem VII}.

\demo{Proof of Proposition 6.3} The formulas for $K^\tail(s,t\mid
z,z')$ are readily obtained from those for $K^\gam(x,y\mid z,z')$,
see Theorem 3.5, using the standard asymptotics of ratios of
gamma--functions, see \cite{Er1, 1.18 (4)}. One also has to keep
in mind the transformation of differentials explained in the proof
of Proposition 6.1.\qed
\enddemo

The statement of Proposition 6.2 also carries over.

\proclaim{Proposition 6.4} In the scaling limit $x=\pm e^{s_0+s}$,
where $s_0\to+\infty$ with
$$
e^{s_0}=O((1-\xi)^{-\epsi}),\qquad 0<\epsi<1-|\Re(z-z')|,
$$
the discrete hypergeometric kernel $K(x,y\mid z,z',\xi)$
converges, as $\xi\nearrow 1$, to the tail kernel $K^\tail(s,t\mid
z,z')$.
\endproclaim
The proof goes along the same lines as that of Proposition 6.2,
and we omit it.

Proposition 6.3 and 6.4 prove the convergence of the correlation
kernels. In many situations it is simpler to establish the
corresponding convergence of $L$-kernels (as usual,
$L=K(1-K)^{-1}$, where $K$ is a correlation kernel). We have
already used the convergence of $L$-kernels, see Theorems 5.5 and
5.6. The following statement shows that the convergence of the
discrete hypergeometric kernel and the gamma kernel to the tail
kernel can be seen on the level of the corresponding $L$-kernels.

\proclaim{Proposition 6.5} In the scaling limits of Propositions
6.3, 6.4, the kernels $L(x,y\mid z,z')$ defined at the end of \S5,
and $L(x,y\mid z,z',\xi)$ defined by \tht{3.2} converge to a
kernel on $\R\times \R$ given by
$$
\align \bullet& \text{    For $x=e^{s_0+s}$, $y=e^{s_0+t}$, the
kernel is identically equal to 0}\\
\bullet&\text{   For $x=e^{s_0+s}$, $y=-e^{s_0+t}${\rm:}     }
\quad\frac{\sqrt{\sin(\pi z)\sin(\pi z')}}{\pi}\cdot
\frac{e^{\frac12(z+z')(s-t)}}{2\cosh(\frac 12(s-t))}\\
\bullet&\text{   For $x=-e^{s_0+s}$, $y=e^{s_0+t}${\rm:}     }
\quad -\frac{\sqrt{\sin(\pi z)\sin(\pi z')}}{\pi}\cdot
\frac{e^{-\frac12(z+z')(s-t)}}{2\cosh(\frac 12(s-t))}\\
\bullet&\text{   For $x=-e^{s_0+s}$, $y=-e^{s_0+t}$, the kernel is
also identically equal to 0.}
\endalign
$$
\endproclaim
\demo{Proof} Direct computation. \qed
\enddemo

Let us denote the kernel defined in Proposition 6.5 by
$L^\tail(s,t)$. It is easy to see that it defines a bounded
operator in $L^2(\R\sqcup \R)$ if and only if $|z+z'|<1$.
Similarly to Theorems 3.4 and 5.5, we have the following claim.

\proclaim{Proposition 6.6 \cite{Ol1, Proposition 4.2}} If
$|z+z'|<1$ then
$$
K^\tail=\frac{L^\tail}{1+L^\tail}.
$$
\endproclaim
\demo{Proof} Let us identify $\L^2(\R\cup\R)$ with $L^2(\R)\oplus
L^2(\R)$. Then we may interpret integral operators in this Hilbert
space as $2\times2$ matrix--valued integral operators on $\R$.
Thus, we may write
$$
L^\tail=\bmatrix L^\tail_{11} & L^\tail_{12}\\
L^\tail_{21} & L^\tail_{22} \endbmatrix\,, \qquad K^\tail=\bmatrix
K^\tail_{11} & K^\tail_{12}\\ K^\tail_{21} & L^\tail_{22}
\endbmatrix\,,
$$
where all the blocks are integral operators in $L^2(\R)$,
$$
L^\tail_{ij}=L^\tail_{ij}(s,t), \qquad
K^\tail_{ij}=K^\tail_{ij}(s,t), \qquad i,j=1,2.
$$
Actually, these integral operators are translationally invariant,
so that we may write
$$
L^\tail_{ij}(s,t)=L^\tail_{ij}(s-t), \qquad
K^\tail_{ij}(s,t)=K^\tail_{ij}(s-t), \qquad i,j=1,2.
$$

Given a function $f(s)$ on $\R$, let $\widehat f(u)$ denote its
Fourier transform,
$$
\wh f(u)=\int_{-\infty}^{+\infty}e^{ius}f(s)ds.
$$
The Fourier transform is an isometry between $L^2(\R, ds)$ and
$L^2(\R, \frac{du}{2\pi})$. By virtue of the translation
invariance, the Fourier images of $L^\tail$ and $K^\tail$ are
operators of multiplication by $2\times2$ matrix--valued functions
in $u$:
$$
\wh L^\tail(u)=\bmatrix \wh L^\tail_{11}(u) & \wh L^\tail_{12}(u)\\
\wh L^\tail_{21}(u) & \wh L^\tail_{22}(u) \endbmatrix\,, \qquad
\wh
K^\tail(u)=\bmatrix \wh K^\tail_{11}(u) & \wh K^\tail_{12}(u)\\
\wh K^\tail_{21}(u) & \wh L^\tail_{22}(u)
\endbmatrix\,,
$$

{}From the explicit expressions for $L$ and $K$ (see Propositions
6.3 and 6.5) it follows that their Fourier images have the form
$$
\wh L^\tail(u)=\bmatrix 0 & c(u)\\
-\overline{c(u)} & 0 \endbmatrix\,, \qquad \wh
K^\tail(u)=\bmatrix a(u) & b(u)\\
-\overline{b(u)} & a(u)
\endbmatrix\,,
$$
where
$$
\gather c(u)=\left\{\frac{\sqrt{\sin(\pi z)\sin(\pi z')}}
\pi\cdot\frac{e^{\frac12(z+z')s}}{2\cosh(\frac
s2)}\right\}^{\wedge}_{s\to u}
\\
a(u)=\left\{\frac{\sin(\pi z)\sin(\pi
z')}{\pi\sin(\pi(z-z'))}\cdot \frac{\sinh(\frac12(z-z')s)}
{2\sinh(\frac s2)}\right\}^{\wedge}_{s\to u}
\\
b(u)=\left\{\frac{\sqrt{\sin(\pi z)\sin(\pi
z')}}{\pi\sin(\pi(z-z'))}\cdot \frac{\sin(\pi z)e^{\frac12(z-z')s}
-\sin(\pi z')e^{-\frac12(z-z')s}}{2\cosh(\frac
s2)}\right\}^{\wedge}_{s\to u}
\endgather
$$

The required Fourier images can be evaluated from the tables, see
formulas 1.9(14) and 3.2(15) in Erdelyi \cite{Er2},
$$
\gather \left\{\frac{\sinh(\frac12(z-z')s)}{\sinh(\frac12
s)}\right\}^{\wedge}_{s\to u}=\frac{2\pi\sin(\pi(z-z'))}{\cos(2\pi
i u)+\cos(\pi(z-z'))}
\\
 \left\{\frac{e^{\frac12(z\pm
z')s}}{2\cosh(\frac s2)}\right\}^{\wedge}_{s\to u}=\frac
\pi{\cos(\pi i u -\frac12\pi(z\pm z'))}
\endgather
$$

{}From these formulas we get explicit expressions
$$
\gather
c(u)=\frac{\sqrt{\sin(\pi z)\sin(\pi z')}}{\cos(\pi i u-\frac12\pi(z+z'))}\\
a(u)=\frac{2\sin(\pi z)\sin(\pi z')}{\cos(2\pi i u)+\cos(\pi(z-z'))}\\
b(u)=2\sqrt{\sin(\pi z)\sin(\pi z')}\; \frac{\cos(\pi i u+\frac12
\pi(z+z'))}{\cos(2\pi i u)+\cos(\pi(z-z'))}
\endgather
$$
Now, the claim of the proposition is equivalent to the relations
$$
a(u)=\frac{c(u)\overline{c(u)}}{1+c(u)\overline{c(u)}}, \qquad
b(u)=\frac{c(u)}{1+c(u)\overline{c(u)}},
$$
which are checked directly from the above expressions. \qed
\enddemo

\head 7. ZW-measures on signatures\endhead

In this section, we replace the set $\Y$ of Young diagrams by the set $\SGN(N)$
of {\it signatures\/} of length $N$. Here $N=1,2,\dots$, and a signature
$\la\in\SGN(N)$ is an $N$--tuple of weakly decreasing integers,
$$
\la=(\la_1,\dots,\la_N), \qquad \la_1\ge\dots\ge\la_N\,.
$$
We will describe a family of probability measures on the sets
$\SGN(N)$ (for more detail, see Olshanski \cite{Ol2} and
Borodin--Olshanski \cite{BO4}). Then we will study the behavior of
the measures as $N\to\infty$, there the limit transition is
similar to the ``second regime'' considered in \S2 above. We show
that the final result is again described in terms of the gamma
kernel.

Our probability measures on $\SGN(N)$ depend on 4 complex parameters
$z,z',w,w'$ and have the form
$$
M_{z,z',w,w'\mid N}(\la)= (\const_N)^{-1}\cdot M'_{z,z',w,w'\mid N}(\la)
$$
where
$$
\gather M'_{z,z',w,w'\mid N}(\la)= \prod_{i=1}^N
\bigg(\frac1{\Gamma(z-\la_i+i)\Gamma(z'-\la_i+i)}\\
\times\frac1{\Gamma(w+N+1+\la_i-i)\Gamma(w'+N+1+\la_i-i)}\bigg)\cdot
(\Dim_N(\la))^2,\\
\Dim_N(\la)=\prod_{1\le i<j\le N} \frac{\la_i-\la_j+j-i}{j-i}\,,
\endgather$$
and
$$
\const_N=\sum_{\la\in\SGN(N)}M'_{z,z',w,w'\mid N}(\la)
$$
is the normalizing constant depending on $z,z',w,w',N$. Under
suitable conditions on the quadruple $(z,z',w,w')$, the measures
$M_{z,z',w,w'\mid N}$ are well defined for all $N$. That is,  the
weights $M'_{z,z',w,w'\mid N}(\la)$ are nonnegative and their sum
over $\la\in\SGN(N)$ is finite. A criterion for that to happen and
for all the weights $M_{z,z',w,w'\mid N}(\la)$ to be strictly
positive is provided below. See \cite{Ol2, \S7} for detailed
explanations.

The measures $M_{z,z',w,w'\mid N}$ can be obtained by a
construction which is quite similar to that described in \S1, with
the finite symmetric group $S_n$ replaced by the compact group
$U(N)$ of unitary $N\times N$ matrices. Let $\mu_N$ be the
normalized Haar measure on $U(N)$, and let $H_N$ be the Hilbert
space of square integrable functions on $U(N)$ (with respect to
$\mu_N$), constant on conjugacy classes. In $H_N$, there is a
distinguished orthonormal basis formed by the irreducible
characters $\chi_\la$ of the group $U(N)$. Here $\la$ ranges over
$\SGN(N)$.

Let $\T$ be the unit circle in $\C$ and $\T^N$ be the product of
$N$ copies of $\T$ (the $N$--dimensional torus). Given a unitary
matrix $U\in U(N)$, we assign to it the unordered $N$--tuple of
its eigenvalues, $(u_1,\dots,u_N)$. Any element of $H_N$ can be
viewed as a function in $(u_1,\dots,u_N)$, that is, as a symmetric
function on the torus $\T^N$. In particular, the irreducible
characters $\chi_\la(U)$ are the (rational) Schur functions
$s_\la(u_1,\dots,u_N)$,
$$
s_\la(u_1,\dots,u_N) =\frac{\det\limits_{1\le i,j\le N}[u_i^{\la_j+N-j}]}
{\det\limits_{1\le i,j\le N}[u_i^{N-j}]}\,.
$$
The whole Hilbert space $H_N$ can be identified with the Hilbert space of
symmetric functions on $\T^N$, square integrable with respect to the measure
$$
\bar\mu_N(du)=\frac1{N!}\, \prod_{1\le i<j\le N}|u_i-u_j|^2 \prod_{i=1}^N
du_i\,,
$$
which is the push--forward of $\mu_N$ under the correspondence
$U\mapsto(u_1,\dots,u_N)$. Here $du_i$ is the normalized invariant measure on
the $i$th copy of $\T$.

Given two complex numbers $z,w$, we define a symmetric function on
$\T^N$ by
$$
f_{z,w\mid N}(u)=\prod_{i=1}^N (1+u_i)^z(1+\bar u_i)^w.
$$
If $\Re(z+w)>-\frac12$ then $f_{z,w\mid N}$ belongs to the space $H_N$. Let
$(z',w')$ be another couple of complex numbers with $\Re(z'+w')>-\frac12$. We
set
$$
M_{z,z',w,w'\mid N}(\la)=\frac{(f_{z,w\mid N},
\chi_\la)(\chi_\la,f_{\overline{w'},\, \overline{z'}\mid N})} {(f_{z,w\mid
N},f_{\overline{w'}, \,\overline{z'}\mid N})}\,,  \qquad \la\in\SGN(N),
$$
where $(\,\cdot\,,\,\cdot\,)$ is the inner product in $H_N$. It
turns out that this definition leads us to the explicit formula
given above. Notice that $\Dim_N\la$ is equal to the value of the
character $\chi_\la$ at $1\in U(N)$ (equivalently, to the value of
the Schur function $s_\la$ at $(1,\dots,1)\in\T^N$).

Similarly to the identification of the Young diagrams with points
in $\{0,1\}^{\Z'}$ described at the beginning of \S2, we identify
a signature $\la=(\la_1,\dots,\la_N)$ with a binary sequence
$\unX(\la)=(\dots,a_{-3/2},a_{-1/2}\,|\,a_{1/2},a_{3/2}\dots)$ by
$$
a_{j}=\cases 1,&
\text{if  } j\in\{\la_i-i+\frac 12\,|\,i=1,\dots,N\},\\
             0,& \text{otherwise}.
\endcases
$$
Note that this identification establishes a one-to-one
correspondence between $\SGN(N)$ and the elements from
$\{0,1\}^{\Z'}$ with exactly $N$ 1's.

In what follows we will assume that neither of the parameters
$z,z',w,w'$ is an integer. This is always the case if we require
the weights of all signatures to be nonzero. Further, the
condition of $M_{z,z',w,w'\mid N}(\la)$ of being positive for all
$\la\in\SGN(N)$ is equivalent to both pairs $(z,z')$ and $(w,w')$
satisfying one of the conditions (i) and (ii) of Proposition 1.8,
and the convergence of the series
$\sum_{\la\in\SGN(N)}M_{z,z',w,w'\mid N}(\la)$ is equivalent to
the inequality
$$
\Re(z+z'+w+w')>-1,
$$
see \cite{Ol2} for proofs. We also assume these conditions to be
satisfied.

The following statement is an analog of Theorem 2.2.

\proclaim{Theorem 7.1 (\cite{BO4, Theorem 7.1})} Let
$\unP_{z,z',w,w'\mid N}$ be the push--forward of the measure
$M_{z,z',w,w'\mid N}$ under the embedding $\la\mapsto \unX(\la)$
of $\SGN(N)$ into $\{0,1\}^{\Z'}$ defined above. Then its
correlations functions have determinantal form
$$
\gathered \rho_m(x_1,\dots,x_m\mid\unP_{z,z',w,w'\mid N})=
\det_{1\le i,j\le m}[\unK(x_i,x_j\mid z,z',w,w'\mid N)],\\
m=1,2,\dots,\quad x_1,\dots, x_m\in\Z',
\endgathered
$$
where the correlation kernel is given by
$$
 \unK(x,y\mid z,z',w,w'\mid N)=\frac
1{h_{N-1}}\,\frac{p_N(x) p_{N-1}(y)-p_{N-1}(x)p_N(y)}
{x-y}\,\sqrt{f(x)f(y)}\,,
$$
$$
\gathered
p_N(x)=\frac{\Gamma(x+w'+N+\frac{1}2)}{\Gamma(x+w'+\frac{1}2)}\,
{}_3F_2\left[\matrix -N,\,z+w',\,z'+w'\\ \s,\,x+w'+\frac{1}2
\endmatrix\Biggl|\,1\,\right],\\
p_{N-1}(x)=\frac{\Gamma(x+w'+N+\frac{1}2)}{\Gamma(x+w'+\frac{1}2+1)}\,
{}_3F_2\left[\matrix -N+1,\,z+w'+1,\,z'+w'+1\\
\s+2,\,x+w'+\frac{1}2+1\endmatrix\Biggl|\,1\,\right],\\
h_{N-1}=\Gamma\left[\matrix N,\,\s+1,\,\s+2\\
\s+N+1,\,z+w+1,\,z+w'+1,\,z'+w+1,\,z'+w'+1\endmatrix\right],\\
f(x)=\frac 1{\Gamma\left(z-x+\frac{1}2\right)\Gamma\left(z'-
x+\frac{1}2\right)\Gamma
\left(w+x+N+\frac{1}2\right)\Gamma\left(w'+x+N+\frac{1}2\right)}\,.
\endgathered
$$
with $\s=z+z'+w+w'$.
\endproclaim
\demo{Notation} The symbol ${}_3F_2$ above stands for the higher
hypergeometric series of type (3,2), see e.g. \cite{Er1, chapter
4} and \cite{Ba}. We also use the notation
$$
\Ga\bmatrix a,\,b,\,\dots\\c,\,d,\,\dots\endbmatrix=
\frac{\Ga(a)\Ga(b)\cdots}{\Ga(c)\Ga(d)\cdots}\,.
$$
\enddemo

\demo{Comments} 1. The functions $p_{N-1}$ and $p_N$ are monic
(i.e., the highest coefficient is equal to 1) orthogonal
polynomials on $\Z'$ of degree $(N-1)$ and $N$, corresponding to
the weight function $f(x)$, and $h_{N-1}=\Vert
p_{N-1}\Vert_{\ell^2(\Z',f)}^2$. The determinantal structure of
the correlation functions with the kernel expressed through
orthogonal polynomials as above is a standard fact from Random
Matrix Theory. Up to the factor $\sqrt{f(x)f(y)}$, the kernel is
the Christoffel--Darboux kernel for the orthogonal polynomials
with weight $f(x)$.

2. If $\s=0$ then the formula for $p_N$ above does not make sense
because it involves a hypergeometric function with a zero lower
index. However, the kernel itself admits an analytic continuation
to the set $\s=0$, see \cite{BO4, \tht{7.3}}.
\enddemo

The next statement is an analog of Theorem 2.3.

\proclaim{Theorem 7.2} The measures $\unP_{z,z',w,w'\mid N}$
weakly converge, as $N\to\infty$, to the probability measure
$\unP_{-z,-z'}^\gam$ on $\{0,1\}^{\Z'}$ defined in Theorem
2.3.\footnote{In fact, Theorem 2.3 provides formulas for the
correlation functions of $\unP_{-z,-z'}^\gam$, which uniquely
define the measure.}
\endproclaim
\demo{Proof} Similarly to the proof of Theorem 2.3, we will show
that the kernel $\unK(x,y\mid z,z',w,w'\mid N)$ has a pointwise
limit as $N\to\infty$ which equals the gamma kernel. We will
assume that $x\ne y$, $z\ne z'$, and $\s\ne 0$. The convergence is
easily extended to these sets by analytic continuation, as is
explained at the end of the proof of Theorem 2.3.

We will use the following transformation formula for ${}_3F_2$
with the unit argument, see Bailey \cite{Ba, 3.2(2)}:
$$
\aligned
{}_3F_2\left[\matrix a,\,b,\,c\\
e,\,f\endmatrix\Bigl|\,1\,\right]= &\Ga\left[\matrix
1-a,\,e,\,f,\,c-b\\
e-b,\,f-b,\,1+b-a,\,c\endmatrix\right] {}_3F_2\left[\matrix
b,\,b-e+1,\,b-f+1\\ 1+b-c,\,1+b-a
\endmatrix\Bigl|\,1\,\right]\\
&+\text{ a similar expression with $b$ and $c$ interchanged}.
\endaligned
$$
If $a\to-\infty$, and $b,c,e,f$ are fixed, the ${}_3F_2$'s in the
right-hand side are equal to $1+O(|a|^{-1})$, and using
$$
\frac{\Gamma(1-a)}{\Gamma(1+b-a)}=(-a)^{-b}(1+O(|a|^{-1})), \quad
\frac{\Gamma(1-a)}{\Gamma(1+c-a)}=(-a)^{-c}(1+O(|a|^{-1})),
$$
we obtain
$$
\multline
{}_3F_2\left[\matrix a,\,b,\,c\\
e,\,f\endmatrix\Bigl|\,1\,\right]=(-a)^{-b}\Ga\left[\matrix
e,\,f,\,c-b\\
e-b,\,f-b,\,c\endmatrix\right](1+O(|a|^{-1}))\\+(-a)^{-c}\Ga\left[\matrix
e,\,f,\,b-c\\
e-c,\,f-c,\,b\endmatrix\right](1+O(|a|^{-1})).
\endmultline
$$
Applying this estimate to $p_N$ and $p_{N-1}$ with $a=-N$ and
$-N+1$, respectively, we get
$$
\gathered
p_N(x)=\Bigl(N^{-z-w'}\Gamma\left[\matrix \s,\,z'-z\\
z'+w,\,-z+x+\frac 12,\,z'+w'\endmatrix\right]
\left(1+O(\tfrac 1N)\right)\\
+\text{ a similar expression with $z$ and $z'$ interchanged
}\Bigr)\cdot\Gamma\left(x+w'+N+\tfrac
12\right).\\
p_{N-1}(x)=\Bigl(N^{-z-w'-1}\Gamma\left[\matrix \s+2,\,z'-z\\
z'+w+1,\,-z+x+\frac 12,\,z'+w'+1\endmatrix\right]
\left(1+O(\tfrac 1N)\right)\\
+\text{ a similar expression with $z$ and $z'$ interchanged
}\Bigr)\cdot \Gamma\left(x+w'+N+\tfrac 12\right).
\endgathered
$$
As we substitute these formulas into the expression $p_N(x)
p_{N-1}(y)-p_{N-1}(x)p_N(y)$, we see that the part coming from
$O(\frac 1N)$ is equal to
$$
\Gamma\left(x+w'+N+\tfrac 12\right)\Gamma\left(y+w'+N+\tfrac
12\right)N^{-z-z'-2w'-1}\cdot o(1),
$$
due to the fact that $|\Re(z-z')|<1$ and $N^{\pm (z-z')}O(\frac
1N)=o(1)$ as $N\to\infty$. Furthermore, four of the remaining
eight terms cancel out, and we get (using the relation
$\Gamma(s+1)=s\Gamma(s)$ a few times)
$$
\gathered p_N(x)p_{N-1}(y)-p_{N-1}(x)p_N(y)=
\Gamma\left(x+w'+N+\tfrac 12\right)\Gamma\left(y+w'+N+\tfrac 12\right) \\
\times N^{-z-z'-2w'-1}\Gamma
\left[\matrix \s,\,\s+2,\,z'-z,\,z-
z'\\z+w,\,z'+w,\,z+w',\,z'+w'\endmatrix\right]\\
\times\Biggl(\frac1{\Gamma(-z+x+\frac 12)\Gamma(-z'+y+\frac
12)}\left(\frac 1{(z+w)(z+w')}-\frac 1{(z'+w)(z'+w')}\right)\\+
\frac1{\Gamma(-z'+x+\frac 12)\Gamma(-z+y+\frac 12)}\left(\frac
1{(z'+w)(z'+w')}-\frac 1{(z+w)(z+w')}\right)+o(1)\Biggr).
\endgathered
$$
Simplifying
$$
\frac 1{(z+w)(z+w')}-\frac
1{(z'+w)(z'+w')}=\frac{(z'-z)\,\s}{(z+w)(z'+w)(z+w')(z'+w')}\,
$$
and using the formula for $h_{N-1}$, we see that
$$
\gathered
\frac{p_N(x)p_{N-1}(y)-p_{N-1}(x)p_N(y)}{h_{N-1}(x-y)}=
\frac{\Gamma(\s+N+1)}{\Ga(N)}\cdot((z'-z)\Gamma(z'-z)\Gamma(z-z'))
\\ \times \,\Gamma\left(x+w'+N+\tfrac 12\right)
\Gamma\left(y+w'+N+\tfrac 12\right)N^{-z-z'-2w'-1}(1+o(1))
\\
\times \left(\frac1{\Gamma(-z+x+\frac 12)\Gamma(-z'+y+\frac
12)}-\frac1{\Gamma(-z'+x+\frac 12)\Gamma(-z+y+\frac
12)}\right)\frac 1{x-y}\,.
\endgathered
$$
Since $\Gamma(\s+N+1)/\Ga(N)\sim N^{\s+1}$ and
$(z'-z)\Ga(z'-z)\Ga(z-z')=\pi/\sin(\pi(z-z'))$, we see that the
above expression equals
$$
\gathered \frac{\pi}{\sin(\pi(z-z'))}\,\Gamma\left(x+w'+N+\tfrac
12\right)\Gamma\left(y+w'+N+\tfrac
12\right)N^{w-w'}(1+o(1))\\
\times \left(\frac1{\Gamma(-z+x+\frac 12)\Gamma(-z'+y+\frac
12)}-\frac1{\Gamma(-z'+x+\frac 12)\Gamma(-z+y+\frac
12)}\right)\frac 1{x-y}\,.
\endgathered
$$
It remains to multiply this expression by $\sqrt{f(x)f(y)}$ and
take the limit $N\to\infty$. The weight function $f(x)$ consists
of four gamma--factors, two of which do not depend on $N$ while
the two others do. Taking the factors in $\sqrt{f(x)f(y)}$ which
are independent on $N$, we obtain
$$
\gathered \frac1{\sqrt{\Ga(z-x+\frac12)\Ga(z'-x+\frac12)
\Ga(z-y+\frac12)\Ga(z'-y+\frac12)}}\\
\times \left(\frac1{\Gamma(-z+x+\frac 12)\Gamma(-z'+y+\frac
12)}-\frac1{\Gamma(-z'+x+\frac 12)\Gamma(-z+y+\frac 12)}\right)
\\=\frac{\sin(\pi z)\sin(\pi z')}{\pi^2\sqrt{\Ga(-z+x+\frac12)
\Ga(-z'+x+\frac12) \Ga(-z+y+\frac12)\Ga(-z'+y+\frac12)}}
\\ \times \left(
\Gamma(-z'+x+\tfrac 12)\Gamma(-z+y+\tfrac 12)-\Gamma(-z+x+\tfrac
12)\Gamma(-z'+y+\tfrac 12)
 \right).
\endgathered
$$
Here we used the fact that due to our restrictions on $(z,z')$,
the product $\sin(\pi z)\sin(\pi z')$ is always positive, so we
can pull it out of the square root.

As for the gamma-factors in $\sqrt{f(x)f(y)}$ that do depend on
$N$, we get
$$
\gathered \frac {\Gamma\left(x+w'+N+\tfrac
12\right)\Gamma\left(y+w'+N+\tfrac
12\right)N^{w-w'}}{\sqrt{\Ga(w+x+N+\frac 12)\Ga(w'+x+N+\frac
12)\Ga(w+y+N+\frac 12)\Ga(w'+y+N+\frac 12)}}\\
=1+O(\tfrac 1N).
\endgathered
$$
Thus, gathering all pieces together, we see that as $N\to\infty$
we have the estimate
$$
\gathered
 \unK(x,y\mid z,z',w,w'\mid N)=
\frac{p_N(x)p_{N-1}(y)-p_{N-1}(x)p_N(y)}{h_{N-1}(x-y)}\sqrt{f(x)f(y)}
\\= \unK^\gam(x,y\mid -z,-z')\cdot (1+o(1)). \qed
\endgathered
$$
\enddemo

\example{Remark 7.3} Observe that the measure $M_{z,z',w,w'\mid
N}$ has the following symmetry property:
$$
M_{z,z',w,w'\mid N}(\la_1,\dots,\la_N)= M_{w,w',z,z'\mid
N}(-\la_N,\dots,-\la_1).
$$
Hence, the measure $\unP_{z,z',w,w'\mid N}$ on $\{0,1\}^{\Z'}$ is
invariant with respect to the simultaneous switch
$(z,z')\longleftrightarrow (w,w')$ of the parameters and the
involution
$$
\{0,1\}^{\Z'}\to \{0,1\}^{\Z'},\qquad (a_j)_{j\in\Z'}\mapsto
(\widehat a_j=a_{-N-j})_{j\in\Z'}.
$$
This means that Theorem 7.2 also implies the following claim:
Embed $\SGN(N)$ into $\{0,1\}^{\Z'}=\{(\dots,a_{-3/2},a_{-1/2}\mid
a_{1/2},a_{3/2},\dots)\}$ by
$$
a_j=\cases 1,& \text{if  } j\in\{-\la_i-(N+1-i)+\frac 12\mid i=1,\dots,N\},\\
            0,&\text{otherwise}.
            \endcases
$$
Then the push--forwards of the measures $M_{z,z',w,w'\mid N}$
under these embeddings weakly converge to $\unP^\gam_{-w,-w'}$ as
$N\to\infty$.
\endexample

\example{Remark 7.4} It is natural to ask whether the two limit
transitions, the one of Theorem 7.2 and the one described in
Remark 7.3 above, lead to asymptotically independent random
point processes. The answer turns out to be positive, and the
exact statement is as follows.

Consider an embedding of $\SGN(N)$ into $\{0,1\}^{\Z'}\times
\{0,1\}^{\Z'}$ defined by using the map $\la\mapsto \unX(\la)$
(described just before Theorem 7.1) on the first coordinate, and
using the map described in Remark 7.3 on the second coordinate.
Then the push--forwards of the measures $M_{z,z',w,w'\mid N}$
under these embeddings converge, as $N\to\infty$, to the product
measure $\unP^\gam_{-z,-z'}\otimes \unP^\gam_{-w,-w'}$.

The proof follows from the fact that the correlation kernel
$\unK(x,y\mid z,z',w,w'\mid N)$ tends to zero as $N\to\infty$ if
one of the arguments $(x,y)$ is in a finite neighborhood of 0,
while the other one is in a finite neighborhood of $-N$. To
prove such an estimate one uses the symmetry of the polynomials
$p_N(x)$ and $p_{N-1}(x)$ with respect to
$(z,z',w,w',x)\longleftrightarrow (w,w',z,z',N-x)$ (which
follows from the obvious symmetry of the weight function
$f(x)$), and the same estimate of the ${}_3F_2$ series as was
used in the proof of Theorem 7.2 above.

\endexample

Similarly to the discrete hypergeometric kernel, the kernel
$\unK(x,y\mid z,z',w,w'\mid N)$ of Theorem 7.1 also has a second
form. This form corresponds to a representation of the signatures
$\la\in\SGN(N)$ through Frobenius coordinates. Given a signature
$\la\in\SGN(N)$ we view it as a pair of Young diagrams
$(\la^+,\la^-)$: one consists of positive $\la_i$'s and the other
one consists of minus negative $\la_i$'s, zeros can go in either
of the two:
$$
\la=(\la_1^+,\la_2^+,\dots,-\la_2^-,-\la_1^-).
$$
Write the diagrams $\la^+$ and $\la^-$ through their Frobenius
coordinates:
$$
\la^{\pm}=(p_1^{\pm},\dots,p_{d^\pm}^{\pm}\mid
q_1^{\pm},\dots,q_{d^\pm}^\pm).
$$
Now we associate to the signature $\la$ a finite subset
$X(\la)\subset \Z'$ (or, equivalently, an element in
$\{0,1\}^{\Z'}$) as follows:
$$
X(\la)=\{p_i^++\tfrac 12\}\sqcup\{-q_i^+-\tfrac
12\}\sqcup\{-p_j^--N-\tfrac 12\}\sqcup\{q_j^--N+\tfrac 12\},
$$
where $i=1,\dots, d^+$ and $j=1,\dots,d^-$. Then we have the
following analog of Proposition 3.1.

\proclaim{Proposition 7.5 (\cite{BO4, Proposition 4.1})} For any
$\la\in\SGN(N)$, the two finite subsets $\unX(\la)$ and $X(\la)$
are related by
$$
X(\la)=\unX(\la)\triangle \left\{-\tfrac 12,-\tfrac 32, \dots,
-N+\tfrac 12\right\},\quad \unX(\la)=X(\la)\triangle
\left\{-\tfrac 12,-\tfrac 32, \dots, -N+\tfrac 12\right\},
$$
where $\triangle$ denotes the symmetric difference of two sets.
\endproclaim

Note that the notation in \cite{BO4} is slightly different, all
points are shifted to the right by $N/2$ comparing to our notation
here.

Theorem 8.7 of \cite{BO4} proves that the push--forward of the
measure $M_{z,z',w,w'\mid N}$ under the map $\la\mapsto X(\la)$
has determinantal correlation functions and gives explicit
formulas for the kernel. Let us denote the correlation kernel by
$K(x,y\mid z,z',w,w'\mid N)$. (Once again, this kernel is
different from that in \cite{BO4} by the shift $x\mapsto x+\frac
N2$.)

The two correlation kernels, $\unK(x,y\mid z,z',w,w'\mid N)$ of
\cite{BO4, Theorem 7.1} used in Theorems 7.1, 7.2 above, and
$K(x,y\mid z,z',w,w'\mid N)$ are related by a simple transform
similar to that of Theorems 4.2 and 4.4. This fact is explained in
\cite{BO4, Theorem 5.10} in a fairly general framework. Together
with Theorems 4.4 and 7.2, this implies that $K(x,y\mid
z,z',w,w'\mid N)$ converges to $K^\gam(x,y\mid -z,-z')$ as
$N\to\infty$.

Note that, similarly to Theorem 5.3, the kernel $\unK(x,y\mid
z,z',w,w'\mid N)$ represents an  orthogonal projection operator by
the very definition; its range is the $N$-dimensional space
$\operatorname{Span}\{\sqrt{f(x)},x\sqrt{f(x)},\dots,
x^{N-1}\sqrt{f(x)}\}$.

Furthermore, \cite{BO4, Theorem 8.7} shows that $K(x,y\mid
z,z',w,w'\mid N)$ has a rather simple $L$-kernel, $L=K(1-K)^{-1}$,
presented in \cite{BO4, \S6}. (This is an analog of Theorem 3.4
above.) Using the explicit form of this $L$-kernel, it is not hard
to show that a natural analog of Proposition 6.5 holds true.
However, this is not enough to ensure the convergence of the
correlation kernels for all admissible values of parameters (the
reason being the unboundedness of the limit $L$-kernels for
$|z+z'|\ge1 )$. Hence, it is of interest to evaluate the
asymptotic behavior of the correlation kernel directly, so we
state this as

\proclaim{Problem 7.6 (cf. Proposition 6.4)} Show that in the
scaling limit $x=\pm e^{s_0+s}$, where $s_0\to +\infty$ with
$$
e^{s_0}=O(N^\epsi), \qquad 0<\epsi<1-|\Re(z-z')|,
$$
the kernel $K(x,y\mid z,z',w,w'\mid N)$ converges to the tail
kernel $K^\tail(s,t\mid -z,-z')$.
\endproclaim

Due to the symmetry explained in Remark 7.3, solving this problem
will also imply the convergence of $K(x,y\mid z,z',w,w'\mid N)$ in
the scaling limit $x=-N\mp e^{s_0+s}$ to the tail kernel
$K^\tail(s,t\mid -w,-w')$.

\head 8. Z-measures on nonnegative signatures
\endhead

In this section, we deal with the subset $\SGN^+(N)\subset\SGN(N)$
formed by the signatures $\la\in\SGN(N)$ with $\la_N\ge0$.
Elements of $\SGN^+(N)$ may be called {\it nonnegative
signatures\/} of length $N$. We will consider a family of
probability measures on $\SGN^+(N)$ depending on parameters
$z,z',a,b$, where $(z,z')$ is a couple of complex numbers
satisfying suitable conditions, and $a, b$ are real numbers such
that $a>-1$, $b>-1$. It is convenient to denote
$$
\epsi=\frac{a+b+1}2\,.
$$

We set
$$
M_{z,z',a,b\mid N}(\la)=(\const_N)^{-1}\cdot M'_{z,z',a,b\mid N}(\la), \qquad
\la\in\SGN^+(N),
$$
where
$$
\gather M'_{z,z',a,b\mid N}(\la)= \prod_{i=1}^N
\Bigg(\frac{(N+\epsi+\la_i-i)\Gamma(N+2\epsi+\la_i-i)
\Gamma(N+a+1+\la_i-i)}{\Gamma(N+b+1+\la_i-i)\Gamma(N+1+\la_i-i)}\\
\times\frac1
{\Gamma(z-\la_i+i)\Gamma(z'-\la_i+i)
\Gamma(z+2N+2\epsi+\la_i-i)\Gamma(z'+2N+2\epsi+\la_i-i)}\Bigg)\\
\times\prod_{1\le i<j\le
N}\left((N+\la_i-i+\epsi)^2-(N+\la_j-j+\epsi)^2\right)^2
\endgather
$$
and
$$
\const_N=\sum_{\la\in\SGN^+(N)}M'_{z,z',a,b\mid N}(\la).
$$

One sufficient condition ensuring the existence of the probability
measures for all $N$ is $z'=\bar z$, $\Re z>-\frac{1+b}2$.

Once again, the above formula can be obtained following the same
general scheme. As the Hilbert space $H_N$ we now take the space
of symmetric functions on the $N$--dimensional cube $[-1,1]^N$,
square integrable with respect to the measure
$$
\bar\mu_N(dx)=\prod_{1\le i<j\le N}(x_i-x_j)^2\cdot \prod_{i=1}^N
(1-x_i)^a(1+x_i)^b \cdot dx_1\dots dx_n\,.
$$

A distinguished orthogonal basis in $H_N$ is formed by the multivariate Jacobi
polynomials
$$
P^{a,b}_{\la\mid N}(x_1,\dots,x_N)=\frac{\det\limits_{1\le i,j\le
N}[P^{a,b}_{\la_j+N-j}(x_i)]} {\det\limits_{1\le i,j\le
N}[P^{a,b}_{N-j}(x_i)]}\,,
$$
where $P^{a,b}_{m}(y)$ are the classical Jacobi polynomials, orthogonal on the
segment $-1\le y\le 1$ with the weight function $(1-y)^a(1+y)^b$. Let us set
$$
\chi^{a,b}_\la(x_1,\dots,x_N) =\frac{P^{a,b}_{\la\mid N}(x_1,\dots,x_N)} {\Vert
P^{a,b}_{\la\mid N}\Vert}, \qquad \la\in\SGN^+(N),
$$
where $\Vert\,\cdot\,\Vert$ stands for the norm in $H_N$. The normalized
polynomials $\chi^{a,b}_\la$ form an orthonormal basis in $H_N$.

We define
$$
f_{z\mid N}(x_1,\dots,x_N)=\prod_{i=1}^N(1+x_i)^z.
$$
Then the formula for the measure is obtained from the expression
$$
M_{z,z',a,b\mid N}(\la)=\frac{(f_{z\mid N},
\chi^{a,b}_\la)(\chi^{a,b}_\la,f_{\overline{z'}\,\mid N})} {(f_{z\mid
N},f_{\overline{z'}\,\mid N})}\,,  \qquad \la\in\SGN^+(N).
$$

Notice that for special values of $(a,b)$, the multivariate Jacobi
polynomials $P^{a,b}_{\la\mid N}$, suitably renormalized, can be
interpreted as the irreducible characters of the compact classical
groups $O(2N+1)$, $Sp(2N)$, $O(2N)$, or as indecomposable
spherical functions on the complex Grassmannians
$U(2N+k)/U(N+k)\times U(N)$.  See, e.g., Okounkov--Olshanski
\cite{OkOl}, Berezin--Karpelevich \cite{BK}.

Let us take the same embedding $\SGN(N)$ into $\{0,1\}^{\Z'}$
(which is identified with subsets of $\Z'$) as we took in \S7:
$\la\mapsto \unX(\la)=\{\la_i-i+\frac12\}_{i=1}^N$. Denote by
$\unP_{z,z',a,b\mid N}$ the push--forward of the measure
$M_{z,z',a,b\mid N}$ under this embedding. Standard tools of
Random Matrix Theory provide us with the following claim, cf.
Theorem 2.2 and Theorem 7.1.

\proclaim{Proposition 8.1} The correlation functions of the
measure $\unP_{z,z',a,b\mid N}$ have determinantal form
$$
\gathered \rho_m(x_1,\dots,x_m\mid \unP_{z,z',a,b\mid
N})=\det_{1\le i,j\le m}[\unK(x_i,x_j\mid z,z',a,b\mid N)],
\\ m=1,2,\dots,\quad x_1,\dots,x_m\in\Z',
\endgathered
$$
where the correlation kernel  has the form
$$
\unK(x,y\mid z,z',a,b\mid N)=\frac{q_N(\widehat x^2)q_{N-1}
(\widehat y^2)- q_{N-1}(\widehat x^2)q_{N}(\widehat
y^2)}{h_{N-1}\cdot (\widehat x^2-\widehat y^2)}\, \sqrt{g(x)g(y)},
$$
where $\widehat x=N+x+\epsi-\frac 12$, $\widehat
y=N+y+\epsi-\frac12$,  $\{q_l\}_{l\ge 0}$ are monic polynomials,
$\deg q_l=l$, satisfying
$$
\sum_{x\in\Z'}q_k(\widehat x^2)q_l(\widehat
x^2)g(x)=h_k\delta_{kl}
$$
with
$$
\gathered g(x)= \left(N+\epsi+x-\tfrac
12\right)\,\Gamma\left[\matrix N+2\epsi+x-\frac 12,\,N+a+x+\frac
12\\ N+b+x+\frac 12,\,N+x+\frac 12\endmatrix\right]\\ \times
\Gamma\left[\matrix 1\\z-x+\frac 12,\,z'-x+\frac
12,\,z+2N+2\epsi+x-\frac 12,\,z'+2N+2\epsi+x-\frac
12\endmatrix\right].
\endgathered
$$
\endproclaim

Note that the weight function $g(x)$ vanishes when $x\le -(N+\frac
12)$ due to $\Gamma(N+x+\frac 12)$ in the denominator. Also note
that $g(x)$ has a polynomial asymptotics as $x\to+\infty$, namely
$$
g(x)\sim x^{1-4N-2b-2(z+z')}, \qquad x\to+\infty.
$$
We will make the assumption that the $4N$ moment of $g(x)$ is
finite, which will guarantee the existence of $q_l$ up to $l=N$.
This means that $z+z'>1-b$. The measure $\unP_{z,z',a,b\mid N}$
exists under a milder condition of finiteness of the $4(N-1)$
moment of $g(x)$, and our results can be extended to this wider
domain of parameters by analytic continuation. However, we will
not provide a detailed argument in this paper.

The weight function $g(x)$ generalizes that associated with the
classical Racah polynomials, see e.g. \cite{KS}. Namely, if we
assume that $g(x)$ vanishes if $x$ is greater than some fixed
number, which may be achieved by requiring one of the parameters
$z,z'$ to be an integer, then $\{q_l\}$ are exactly the
(normalized) Racah polynomials. However, it is not immediately
obvious how to generalize the Racah polynomials to the nonintegral
values of $z$ and $z'$.

Fortunately, the orthogonal polynomials that we need were recently
computed by Neretin \cite{Ner}. Actually, Neretin considers even
more general situation when the lattice is infinite at both plus
and minus infinity. Let us state his result.

Take arbitrary complex numbers $a_1, a_2, a_3, a_4$ and $\alpha$,
and consider the weight function
$$
w(t\mid a_1,a_2,a_3,a_4;\al)
=\frac{\alpha+t}{\prod_{j=1}^4\Gamma(a_j+\alpha+t)\Gamma(a_j-\alpha-t)}
\,,\qquad t\in\Z.
$$

\proclaim{Proposition 8.2 (\cite{Ner, \S3.4})} The polynomials
$$
\gathered Q_n((t+\al)^2)=\Gamma\left[\matrix
2-a_1-a_2+n,\,2-a_1-a_3+n,\,2-a_1-a_4+n
\\
2-a_1-a_2,\,2-a_1-a_3,\,2-a_1-a_4\endmatrix\right]\\ \times\,
{}_4F_3\left[\matrix
-n,\, n+3-a_1-a_2-a_3-a_4,\,1-a_1+t+\al,\,1-a_1-t-\al\\
2-a_1-a_2,\,2-a_1-a_3,\,2-a_1-a_4\endmatrix\Big|\,1\,\right]
\endgathered
$$
are orthogonal with respect to the weight $w(t)$, and
$$
\gathered H_n=\sum_{t\in\Z}Q_n^2((t+\al)^2)=\frac{\sin(2\pi\al)
\prod_{i,j=1}^4\sin(\pi(a_i+a_j))}{2\pi^6\,\sin(\pi(a_1+a_2+a_3+a_4))}\\
\times
\,\frac{n!\prod_{i,j=1}^4\Gamma(2-a_i-a_j+n)}{(3-a_1-a_2-a_3-a_4+n)
\Gamma(3-a_1-a_2-a_3-a_4+n)}\,.
\endgathered
$$
These statements hold whenever the corresponding series are
convergent.
\endproclaim

Note that the polynomials $Q_n$ are not monic, the highest
coefficient $k_n$ of $Q_n$ is equal to
$$
k_n=(n+3-a_1-a_2-a_3-a_4)_n=\frac{\Gamma(2n+3-a_1-a_2-a_3-a_4)}
{\Gamma(n+3-a_1-a_2-a_3-a_4)}\,.
$$

In what follows we will use Proposition 8.2 to evaluate the
correlation kernel from Proposition 8.1 in terms of hypergeometric
functions in order to prove the following result, cf. Theorems
2.3, 7.2.

\proclaim{Theorem 8.3} The measures $\unP_{z,z',a,b\mid N}$ weakly
converge, as $N\to\infty$, to the probability measure
$\unP_{-z,-z'}^\gam$ on $\{0,1\}^{\Z'}$ defined in Theorem 2.3.
\endproclaim
\demo{Proof} The argument resembles those in the proofs of Theorem
2.3 and 7.1 but it is more technically involved. Once again, we
will compute the pointwise asymptotics of the correlation kernel
and show that it converges to the gamma kernel. The argument below
requires that $x\ne y$ and $z\ne z'$. The result is extended to
these exceptional sets by analytic continuation as explained at
the end of the proof of Theorem 2.3.

First of all, applying the identities
$$
\gathered \Gamma(N+2\epsi+x-\tfrac
12)=\frac{\pi}{\sin(\pi(N+2\epsi+x-\frac
12))\Gamma(-N-2\epsi-x+\frac 32)}\,,\\
 \Gamma(N+a+x+\tfrac
12)=\frac{\pi}{\sin(\pi(N+a+x+\frac 12))\Gamma(-N-a-x+\frac
12)}\,,
\endgathered
$$
we observe that the weight function $g(x)$ of Proposition 8.1 is
proportional to the weight function $w(t\mid a_1,a_2,a_3,a_4;\al)$
with the following identification of parameters:
$$
\gathered t=N+x-\tfrac12,\quad \alpha=\epsi,\\
a_1=1-\epsi,\quad a_2=b+1-\epsi,\quad a_3=z+N+\epsi, \quad
a_4=z'+N+\epsi.
\endgathered
$$
Hence, using this identification and the notation $\widehat
x=N+x+\epsilon-\frac 12$, $\widehat y=N+y+\epsilon-\frac 12$, we
may rewrite the correlation kernel $\unK(x,y\mid z,z',a,b\mid N)$
in the form
$$
\frac{k_{N-1}}{k_{N} H_{N-1}}\,\frac{Q_N(\widehat x^2)Q_{N-1}
(\widehat y^2)- Q_{N-1}(\widehat x^2)Q_{N}(\widehat
y^2)}{(\widehat x^2-\widehat y^2)}\, \sqrt{w(\widehat
x-\epsilon)w(\widehat y-\epsilon)}.
$$
It is the asymptotics of this expression that we are going to
compute.

Our next goal is to transform the ${}_4F_3$ hypergeometric
functions that enter the formulas for $Q_{N-1}$ and $Q_N$ into a
form suitable for the limit transition $N\to\infty$. We will do
this in two steps.

First, we use the formula \cite{Ba, 7.2(1)}:
$$
\multline {}_4F_3\left[\matrix
X,\,Y,\,Z,\,-n\\U,\,V,\,W\endmatrix\Big
|\,1\,\right]=\Ga\left[\matrix V-Z+n,\,W-Z+n,\,V,\,W\\
V-Z,\,W-Z,\,V+n,\,W+n\endmatrix\right] \\ \times
{}_4F_3\left[\matrix
U-X,\,U-Y,\,Z,\,-n\\1-V+Z-n,\,1-W+Z-n,\,U\endmatrix\Big
|\,1\,\right]
\endmultline
$$
which holds if the ${}_4F_3$ series are terminating
($n=1,2,\dots$) and {\it Saalsch\"utzian}, that is, the sum of
upper indices is greater than the sum of the lower indices by one:
$U+V+W=X+Y+Z-n+1$.

Applying this formula to $Q_N$ with $n=N$ and
$$
\gathered X=N+3-a_1-a_2-a_3-a_4=1-(z+z'+N+b),\\ Y=1-a_1-t-\al=
-N-x+\tfrac12\,,\\ Z=1-a_1+t+\al=N+2\epsi+x-\tfrac12=
N+a+b+x+\tfrac 12\,\\
U=2-a_1-a_2=a+1,\\ V=2-a_1-a_3=1-z-N,\qquad W=2-a_1-a_4=1-z'-N,
\endgathered
$$
we obtain
$$
\gathered Q_N(\widehat x^2)=\Gamma\left[\matrix 2N+z+a+b+x+\frac
12,\,
2N+z'+a+b+x+\frac 12,\,N+a+1\\
N+z+a+b+x+\frac 12,\, N+z'+a+b+x+\frac 12,\,a+1
\endmatrix \right]\\
\times {}_4F_3\left[\matrix N+z+z'+a+b,\,N+a+x+\frac
12,\,N+a+b+x+\frac 12,\,-N\\ N+z+a+b+x+\frac12,\,N+z'+a+b+x+\frac
12,\,a+1\endmatrix\Big|\,1\,\right].
\endgathered
$$
Similarly,
$$
\gathered Q_{N-1}(\widehat x^2)=\Gamma\left[\matrix
2N+z+a+b+x+\frac 12,\,
2N+z'+a+b+x+\frac 12,\,N+a\\
N+z+a+b+x+\frac 12+1,\, N+z'+a+b+x+\frac 12+1,\,a+1
\endmatrix \right]\\
\times {}_4F_3\left[\matrix N+z+z'+a+b+1,\,N+a+x+\frac
12,\,N+a+b+x+\frac 12,\,-N+1\\
N+z+a+b+x+\frac12+1,\,N+z'+a+b+x+\frac
12+1,\,a+1\endmatrix\Big|\,1\,\right].
\endgathered
$$
The second transformation formula for ${}_4F_3$ that we are about
to use looks as follows:
$$
\gathered {}_4F_3\left[\matrix
X,\,Y,\,Z,\,-n\\U,\,V,\,W\endmatrix\Big |\,1\,\right]= \Ga\bmatrix
1+X-U,\,1+Y-U,\,1+Z-U,\,1-n-U,\,V,\,V-W\\
V-X,\,V-Y,\,V-Z,\,V+N,\,1-U,\,1-U+W
\endbmatrix\\
\times {}_4F_3\left[\matrix
W-X,\,W-Y,\,W-Z,\,W+n\\
1-U+W,\,1-V+W,\,W
\endmatrix\Big|\,1\,\right]\\
+\text{  a similar expression with $V$ and $W$ interchanged }.
\endgathered
$$
This formula also holds for a terminating Saalsch\"utzian
${}_4F_3$ series, and it can be obtained by successful
applications of \cite{Ba, 7.1(1)} and \cite{Ba, 7.5(3)}.

For $Q_N$, we take $n=N$ and
$$
\gathered X=N+z+z'+a+b,\quad Y=N+a+x+\tfrac 12,\quad
Z=N+a+b+x+\tfrac 12,\\
U=a+1,\quad V=N+z+a+b+x+\tfrac12,\quad W=N+z'+a+b+x+\tfrac12.
\endgathered
$$
Then the transformation formula yields
$$
\gathered Q_N(\widehat x^2)=\Gamma\left[\matrix 2N+z+a+b+x+\frac
12,\,
2N+z'+a+b+x+\frac 12,\,N+a+1\\
N+z+a+b+x+\frac 12,\, N+z'+a+b+x+\frac 12,\,a+1
\endmatrix \right]\\
\times \Ga\bmatrix N+z+z'+b,\,N+x+\tfrac
12,\,N+b+x+\frac12,\,-N-a,\,N+z+a+b+x+\frac12,\,z-z'\\
-z'+x+\frac12,\,z+b,\,z,\,2N+z+a+b+x+\frac
12,\,-a,\,N+z'+b+x+\frac 12
\endbmatrix\\
\times {}_4F_3\left[\matrix
-z+x+\frac 12,\,z'+b,\,z',\,2N+z'+a+b+x+\frac 12\\
N+z'+b+x+\frac 12,\,1-z+z',\,N+z'+a+b+x+\frac 12
\endmatrix\Big|\,1\,\right]\\
+\text{ a similar expression with $z$ and $z'$ interchanged }.
\endgathered
$$
We now see that four gamma--factors cancel out, and also
$$
\Ga\bmatrix N+a+1,\,-N-a\\ a+1,\,-a\endbmatrix=(-1)^N.
$$
Further, we observe that the ${}_4F_3$ factors are of the form
$1+O(\frac 1N)$ as $N\to\infty$. Indeed, this is true about any
$$
{}_4F_3\left[\matrix
X,\,Y,\,Z,\,2N+T\\
N+U,\,V,\,N+W\endmatrix\Big|\,1\,\right]
$$
with finite $X,Y,Z,T,U,V,W$; $V\ne 0,-1,-2,\dots$, as follows from
the series representation of ${}_4F_3$.

Taking this into account we obtain
$$
\gathered Q_N(\widehat x^2)=(-1)^N\Ga(N+z+z'+b)\Ga(N+x+\tfrac
12)\Ga(N+b+x+\tfrac12)\\
\times\Biggl(\Ga\bmatrix 2N+z'+a+b+x+\frac 12,\,z-z'\\
-z'+x+\frac 12,\,z+b,\,z,\,N+z'+a+b+x+\frac12,\,N+z'+b+x+\frac
12\endbmatrix(1+O(\tfrac 1N))\\
+\text{ a similar expression with $z$ and $z'$ interchanged }
\Biggr).
\endgathered
$$
Similarly,
$$
\gathered Q_{N-1}(\widehat
x^2)=(-1)^{N-1}\Ga(N+z+z'+b+1)\Ga(N+x+\tfrac
12)\Ga(N+b+x+\tfrac12)\\
\times\Biggl(\Ga\bmatrix 2N+z'+a+b+x+\frac 12,\,z-z'\\
-z'+x+\frac 12,\,z+b+1,\,z+1,\,N+z'+a+b+x+\frac32,
\,N+z'+b+x+\frac 32\endbmatrix\\ \times(1+O(\tfrac 1N)) +\text{ a
similar expression with $z$ and $z'$ interchanged } \Biggr).
\endgathered
$$
Using a number of times the asymptotic relation
$\Ga(M+c)/\Ga(M)=M^c(1+O(\frac 1M))$, $M\to+\infty$, $c$ is fixed,
we simplify the above expressions to get
$$
\gathered \frac{(-1)^NQ_N(\widehat
x^2)}{\sqrt{\Ga(2N+z+a+b+x+\frac 12)\Ga(2N+z'+a+b+x+\frac 12)}}=
\Ga(N+z+z'+b)\\
\times
\Biggl(\Ga\bmatrix z-z'\\
-z'+x+\frac 12,\,z+b,\,z\endbmatrix(2N)^{\frac{z'-z}2}\,N^{-2z'-a-b}(1+O(\tfrac
1N))\\
+\text{ a similar expression with $z$ and $z'$ interchanged }
\Biggr)
\endgathered
$$
and
$$
\gathered \frac{(-1)^{N-1}Q_{N-1}(\widehat
x^2)}{\sqrt{\Ga(2N+z+a+b+x+\frac 12)\Ga(2N+z'+a+b+x+\frac
12)}}=\Ga(N+z+z'+b+1)\\ \times
\Biggl(\Ga\bmatrix z-z'\\
-z'+x+\frac 12,\,z+b+1,\,z+1\endbmatrix(2N)^{\frac{z'-z}2}\,
N^{-2z'-a-b-2}(1+O(\tfrac 1N))\\
+\text{ a similar expression with $z$ and $z'$ interchanged }
\Biggr).
\endgathered
$$

Now as we compute $Q_N(\widehat x^2)Q_{N-1}(\widehat
y^2)-Q_{N-1}(\widehat x^2)Q_N(\widehat y^2)$ normalized by the
square root of the product of four gamma functions as above, we
observe that the terms involving nonzero powers of $(2N)$ cancel
out leaving a remainder of the form
$$
\Ga(N+z+z'+b)\Ga(N+z+z'+b+1)N^{-2(z+z'+a+b+1)}\cdot o(1),
$$
where we used the fact that $N^{\pm(z-z')}O(\frac 1N)=o(1)$.
Hence, the whole expression equals
$$
\gathered -N^{-2(z+z'+a+b+1)}\Ga\bmatrix
N+z+z'+b,\,N+z+z'+b+1,\,z-z',\,z'-z\\z+1,\,z'+1,\,z+b+1,\,z'+b+1
\endbmatrix
\\ \times \Biggl(\frac{z(z+b)-z'(z'+b)}{\Gamma(-z'+x+\frac12)
\Gamma(-z+y+\frac12)}+\frac{z'(z'+b)-z(z+b)}{\Gamma(-z+x+\frac12)
\Gamma(-z'+y+\frac12)}+o(1) \Biggr).
\endgathered
$$
Simplifying, we obtain
$$
\gathered N^{-2(z+z'+a+b+1)}\Ga\bmatrix
N+z+z'+b,\,N+z+z'+b+1\\z+1,\,z'+1,\,z+b+1,\,z'+b+1
\endbmatrix(1+o(1))
\\ \times \frac{\pi(z+z'+b)}{\sin(\pi(z-z'))}\Biggl(
\frac{1}{\Gamma(-z'+x+\frac12)
\Gamma(-z+y+\frac12)}-\frac{1}{\Gamma(-z+x+\frac12)
\Gamma(-z'+y+\frac12)}\Biggr).
\endgathered
$$
To complete the computation of the asymptotics of the correlation
kernel, it remains to take care of the factors
$$
\frac{k_{N-1}\sqrt{w(\widehat x-\epsi)w(\widehat
y-\epsi)}}{k_NH_N(\widehat x^2-\widehat y^2)}\,.
$$
We see that
$$
\gathered \frac{k_{N-1}}{k_N}=\Ga\bmatrix 2N+1-a_1-a_2-a_3-a_4,\,
N+3-a_1-a_2-a_3-a_4\\
N+2-a_1-a_2-a_3-a_4,\,2N+3-a_1-a_2-a_3-a_4\endbmatrix\\
=\Ga\bmatrix -1-z-z'-b,\,1-z-z'-b-N\\
-z-z'-b-N,\,1-z-z'-b\endbmatrix=-(z+z'+b)(z+z'+b+1)N(1+O(\tfrac
1N))
\endgathered
$$
and (assuming $x\ne y$)
$$
\frac 1{\widehat x^2-\widehat y^2}=\frac
1{(N+x+\epsi-\frac12)^2-(N+y+\epsi-\frac 12)^2}=\frac
1{2N(x-y+o(1))}\,.
$$
Further, let us consider the factor $\sqrt{w(\widehat
x-\epsi)w(\widehat y-\epsi)}$. Observe that out of the eight
gamma-functions that enter the expression
$$
\gathered w(\widehat x-\epsi)=\frac{N+2\epsi+x-\frac
12}{\Ga(-N-a-b-x+\frac 12)\Ga(-a-N-x+\frac 12)\Ga(N+x+\frac
12)\Ga(N+b+x+\frac 12)}\\
\times\,\frac1{\Ga(z-x+\frac 12)\Ga(z'-x+\frac
12)\Ga(2N+z+a+b+x+\frac 12)\Ga(2N+z+a+b+x+\frac 12)}
\endgathered
$$
we have already used the last two to normalize $Q_N$ and $Q_{N-1}$
above. The remaining contribution of $\sqrt{w(\widehat
x-\epsi)w(\widehat y-\epsi)}$ equals
$$
\frac{|\sin(\pi(a+b))\sin(\pi a)|}{\pi^2}\frac {N^{2a+1}(1+O(\frac
1N))}{\sqrt{\Ga(z-x+\frac 12)\Ga(z'-x+\frac 12)\Ga(z-y+\frac
12)\Ga(z'-y+\frac 12)}}\,.
$$
Finally, using the formula of Proposition 8.2, we obtain, using
the periodicity of sine several times,
$$
\gathered
H_{N-1}=\pm\frac{\sin(\pi z)\sin(\pi
z')\sin(\pi(a+b))\sin(\pi
a)\sin(\pi(z+z'+a+b))}{2\pi^4 \sin(\pi(z+z'+b))
(-z-z'-b-1)}\\
\times \frac{\sin(\pi(z+b))\sin(\pi(z'+b))}{\pi^2} \Ga\bmatrix
-z,-z',N,N+a,-N-z-z'-a-b,-z-b,-z'-b\\ -N-z-z'-b\endbmatrix.
\endgathered
$$
Simplifying and using the fact that $H_{N-1}$ must be positive, we
obtain
$$
H_{N-1}=\frac{|\sin(\pi(a+b))\sin(\pi a)|}{2\pi^2} \Ga\bmatrix
N,N+a\\1+z,1+z',1+z+b,1+z'+b\endbmatrix N^{-a}(1+O(\tfrac 1N)).
$$
Gathering all the pieces together, we obtain that the correlation
kernel $\unK(x,y\mid z,z',a,b\mid N)$, up to the factor
$(1+o(1))$, is equal to
$$
\gathered N^{-2(z+z'+a+b+1)}\Ga\bmatrix
N+z+z'+b,\,N+z+z'+b+1\\z+1,\,z'+1,\,z+b+1,\,z'+b+1
\endbmatrix
\\ \times \frac{\pi(z+z'+b)}{\sin(\pi(z-z'))}\Biggl(
\frac{1}{\Gamma(-z'+x+\frac12)
\Gamma(-z+y+\frac12)}-\frac{1}{\Gamma(-z+x+\frac12)
\Gamma(-z'+y+\frac12)}\Biggr)\\
\times (-1)(z+z'+b)(z+z'+b+1)N\cdot\frac 1{2N(x-y)}\\
\times \frac{|\sin(\pi(a+b))\sin(\pi a)|}{\pi^2}\frac
{N^{2a+1}}{\sqrt{\Ga(z-x+\frac 12)\Ga(z'-x+\frac 12)\Ga(z-y+\frac
12)\Ga(z'-y+\frac 12)}}\\
\times \left(\frac{|\sin(\pi(a+b))\sin(\pi a)|}{2\pi^2}
\Ga\bmatrix N,N+a\\1+z,1+z',1+z+b,1+z'+b\endbmatrix
N^{-a}\right)^{-1}
\endgathered
$$
which, thanks to the asymptotic relation
$$
\Ga\bmatrix N+z+z'+b,\,N+z+z'+b+1\\ N,\, N+a\endbmatrix=
N^{2(z+z'+b)+1-a}(1+O(\tfrac 1N)),
$$
is readily seen to be asymptotically equal to $\unK^\gam
(x,y|-z,-z').$ \qed
\enddemo

Similarly to the discrete hypergeometric kernel and the ${}_3F_2$
kernel of \S7, the correlation kernel $\unK(x,y\mid z,z',a,b\mid
N)$ also has a second form $K(x,y\mid z,z',a,b\mid N)$ related to
representing signatures in terms of the Frobenius coordinates.
Using \cite{BO4, Theorem 5.10}, it is easy to show that Theorem
8.3 proved above also implies the convergence of the second form
$K(x,y\mid z,z',a,b\mid N)$ to the second form of the gamma kernel
$K^\gam(x,y\mid -z,-z')$.

One can also compute the $L$-kernel, $L=K(1-K)^{-1}$, and consider
the tail scaling limit of the correlation kernels and the
$L$-kernel, but we will postpone the discussion of these issues
until a later publication.

\vskip 2 true cm

{\smc A.~Borodin}: Mathematics 253-37, Caltech, Pasadena, CA
91125, U.S.A.,

\medskip

and  Dobrushin Mathematics Laboratory, Institute for Information
Transmission Problems, Moscow, RUSSIA.

\medskip

E-mail address: {\tt borodin\@caltech.edu}

\bigskip

{\smc G.~Olshanski}: Dobrushin Mathematics Laboratory, Institute
for Information Transmission Problems, Bolshoy Karetny 19,
127994 Moscow GSP-4, RUSSIA.

\medskip

E-mail address: {\tt olsh\@online.ru}

\enddocument
\end